\documentclass[12pt]{article}
\input{epsf}
\newcommand{\postscriptscale}[2]
{\setlength{\epsfxsize}{#2\hsize} \centerline{\epsfbox{#1}}}

\makeatletter
\def\Ddots{\mathinner{\mkern1mu\raise\p@
\vbox{\kern7\p@\hbox{.}}\mkern2mu
\raise4\p@\hbox{.}\mkern2mu\raise7\p@\hbox{.}\mkern1mu}}
\makeatother
\font\bm=cmmib10 at 10pt
\font\bms=cmmib10 at 7pt \textfont9=\bm \scriptfont9=\bms
\usepackage{amsfonts}
\usepackage{multirow}
\mathchardef\balpha= "790B
\mathchardef\bbeta= "790C
\mathchardef\bTheta= "7902
\mathchardef\bzeta= "7910
\mathchardef\bOmega= "790A
\mathchardef\bGamma= "7900
\mathchardef\bDelta= "7901
\mathchardef\bPhi= "7908
\mathchardef\bphi= "791E
\mathchardef\bomega= "7921
\mathchardef\bxi= "7918
\mathchardef\bet= "7911
\mathchardef\brho= "791A
\mathchardef\btau= "791C
\mathchardef\bmu= "7916
\mathchardef\bvarpi= "7924

\def \lvec{(\kern-.26em(}
\def\pmb#1{\setbox0=\hbox{#1}
\kern-.025em\copy0\kern-\wd0
\kern.05em\copy0\kern-\wd0
\kern-.025em\raise.0433em\box0 }
\mathchardef\btheta= "7912
\usepackage{amsmath}
\usepackage{amssymb}
\usepackage{authblk}
\usepackage{url}
\providecommand{\keywords}[1]{\textbf{\textit{Keywords:  }} #1}

\begin{document}
\title{2n-Stream Conservative Scattering}
\author[1]{ W. A. van Wijngaarden}
\author[2] {W. Happer}
\affil[1]{Department of Physics and Astronomy, York University, Canada}
\affil[2]{Department of Physics, Princeton University, USA}
\renewcommand\Affilfont{\itshape\small}
\date{\today}
\maketitle
\begin{abstract}
\noindent We show how to use matrix methods of quantum mechanics to efficiently and accurately calculate axially symmetric  radiation transfer in clouds, with conservative scattering of arbitrary anisotropy. Analyses of conservative scattering, where the single scattering albedo is $\tilde\omega =1$ and no energy is exchanged between the radiation and scatterers, began with work by Schwarzschild, Milne, Eddington and others\cite{Frisch} on radiative transfer in stars. There the scattering is isotropic or nearly so. It has been difficult to extend traditional methods to highly anisotropic scattering, like  that of sunlight in Earth's clouds. The $2n$-stream method described here is a practical way to handle highly anisotropic, conservative scattering. The basic ideas of the $2n$-stream method are an extension of Wick's seminal work on transport of thermal neutrons by 
isotropic scattering\cite{Wick} to scattering with arbitrary anisotropy.  How to do this for finite absorption and $\tilde\omega<1$ was described in a previous paper\cite{WH1}. But those methods fail for conservative scattering, when $\tilde \omega = 1$. Here we show that minor modifications to the fundamental $2n$-scattering theory for $\tilde\omega <1$ make it suitable for $\tilde\omega = 1$.   
\end{abstract}
\newpage
\section{Introduction \label{ps}}
 In a previous paper, {\it 2n-Stream Radiative Transfer}\cite{WH1} we outlined how to use matrix methods of quantum mechanics to accurately and efficiently analyze radiative transfer with highly anisotropic scattering, like Earth's cloud particulates for sunlight.  To facilitate subsequent discussions, we will refer to this paper as WH and to equation ($x$) in it as (WH-$x$).  

The theory outlined in WH works as long as there is some absorption and the single-scattering albedo $\tilde\omega$ is less than 1.  Some absorption  and emission is always present in nature, so the limitation of WH to $\tilde \omega<1$, is consistent with reality.
However, in many important situations, the absorption can be small compared to elastic scattering. The limit of no absorption at all and unit single-scattering albedo, $\tilde\omega=1$, is called conservative scattering\cite{Chandrasekhar}.
Examples of nearly conservative scattering are Thomson scattering of radiation by free electrons in stars or Mie scattering of sunlight in clouds. There is a large body of theoretical literature focused on  conservative scattering\cite{Frisch,Chandrasekhar}. This paper shows how to extend  the simple and computationally efficient methods of WH to the limiting case of conservative scattering.
\section{Vector notation}
 As in WH, we use Dirac notation and we assume axially symmetric radiation.  Then the time-independent, steady-state version of the  equation of transfer (WH-58) becomes
\begin{equation}
\left(\hat\mu\frac{d}{d\tau}+\hat \eta\right) |I (\tau)\}=B(\tau)\hat\eta|0)
\label{in2}
\end{equation}
The abstract, $2n\times 1$ vector $|I(\tau)\}$ approximates  the monochromatic intensity, $I = I(\mu,\tau)$,  at the vertical optical depth $\tau$ above the bottom of a cloud, and with  direction cosine  $\mu$ with respect to the zenith. The $2n\times 2n$ direction-cosine matrix is denoted by the symbol $\hat\mu$ and  the $2n\times 2n$ decay-efficiency matrix  is denoted by $\hat\eta$. We will discuss both $\hat\mu$ and $\hat \eta$ in more detail below. The right side of (\ref{in2}) describes the isotropic emission of radiation by scattering particles. The Planck intensity, $B(\tau)$, depends on the radiation frequency and on the local temperature, $T(\tau)$, as shown in (WH-7).  The symbol $|0)$ that multiplies $B(\tau)\hat\eta$ is the  monopole basis vector $|l)$ of (WH-35) for $l=0$.

The right {\it multipole basis vector}, $|l)$ and its conjugate left basis vector 
$\lvec l|$, satisfy the orthogonality relation of (WH-32) and the completeness relation of (WH-33)
\begin{eqnarray}
\lvec l|l')=\delta_{ll'},\quad\hbox{and}\quad\sum_{l=0}^{2n-1}|l)\lvec l|=\hat 1.\label{in12cc}
\end{eqnarray}
Here $\hat 1$ denotes a $2n\times 2n$ unit matrix, $|l)$ can be thought of as  a $2n\times 1$ column  vector and $\lvec  l|$ as a $1\times 2n$ row vector. Using (\ref{in12cc}), 
one can write $|I\}$ in terms of the $2n$ multipole amplitudes $\lvec l|I(\tau)\}$,
\begin{equation}
|I (\tau)\}=\sum_{l=0}^{2n-1}|l)\lvec l|I(\tau)\},\label{in4}
\end{equation}
where
\begin{equation}
\lvec l|I(\tau)\}=I_l(\tau)=\frac{1}{2}\int_{-1}^1 d\mu\,P_l(\mu)I(\mu,\tau).\label{in5}
\end{equation}
As shown in (WH-31) and (WH-35), the projections of the left and right multipole basis vectors  $\lvec l|$ and $|l)$ onto the continuous, right and left $\mu$-space basis vectors $|\mu\rangle$ and $\langle \mu|$ are 
\begin{equation}
 \lvec l|\mu\rangle =\frac{1}{2}P_l(\mu),
\label {vc14}
\end{equation}
and
\begin{equation}
\langle\mu|l) = P_{l}(\mu)(2l+1).
\label {vc22}
\end{equation}

Instead of expanding the intensity  vector $|I(\tau)\}$ on $2n$ multipole basis vectors $|l)$ and $\lvec l|$, as in (\ref{in4}), we can expand $|I(\tau)\}$ on the $2n$ {\it stream basis vectors}, $|\mu_i)$  and $\lvec \mu_i|$, the  right and left eigenvectors of the direction cosine matrix $\hat\mu$, defined by (WH-105) as
\begin{equation}
\hat\mu|\mu_i)=\mu_i|\mu_i)\quad \hbox{and}\quad \lvec \mu_i|\hat\mu=\lvec\mu_i|\mu_i.\label{in6}
\end{equation}
In analogy to (\ref{in12cc}) the right and left stream basis vectors $|\mu_i)$ and $\lvec \mu_i|$ satisfy the orthogonality relation of (WH-86) and the completeness relation of (WH-87)
\begin{eqnarray}
\lvec \mu_i|\mu_{i'})=\delta_{ii'},\quad\hbox{and}\quad\sum_{i=1}^{2n}|\mu_i)\lvec \mu_i|=\hat 1, \label{in12c}
\end{eqnarray}
In analogy to (\ref{in4}) we can use
the completeness relation of (\ref{in12c})
to expand the intensity vector as
\begin{equation}
|I (\tau)\}=\sum_{i=1}^{2n}|\mu_i)\lvec \mu_i|I(\tau)\}.\label{in6aa}
\end{equation}
The coefficient $\lvec \mu_i|I(\tau)\}$ of (\ref{in6}) is the product of the {\it Gauss-Legendre weight}, $w_i$, and the  value, $I(\mu_i,\tau)$, of the intensity at the {\it Gauss-Legendre direction cosine}, $\mu_i$.
\begin{equation}
\lvec \mu_i|I(\tau)\}=w_i I(\mu_i,\tau)=\int_{-1}^1 d\mu \lvec \mu_i|\mu\rangle I(\mu,\tau).\label{in7}
\end{equation}
 Experiments are normally designed to measure {\it unweighted intensities}, $I(\mu_i,\tau)=w_i^{-1}\lvec\mu_i|I(\tau)\}$, for which a common unit is W/(m$^{2}$ cm$^{-1}$ sr)  or watts per square meter $\times$ wave number $\times$ steradian.  {\it Weighted intensities}, $\lvec\mu_i|I(\tau)\}$, are the natural quanties for mathematical calculations. They are the elements of the $2n\times 1$ intensity vector, $|I(\tau)\}$ in $\mu_i$-space.

According to  (WH-17) and (WH-36), the $2n\times 2n$ elements of the direction-cosine matrix $\hat \mu$ in multipole space are
\begin{eqnarray}
\lvec l|\hat\mu|l')&=&\frac{(l)\delta_{l',l-1}+(l+1)\delta_{l',l+1}}{2l+1}\nonumber\\
	&=&\left[\begin{array}{lllllll} 0&1/1&0&0&0&0&\cdots \\ 1/3&0&2/3&0&0&0&\cdots \\ 
 0&2/5&0&3/5&0&0&\cdots \\   0&0&3/7&0&4/7&0&\cdots \\ 
0&0&0&4/9&0&5/9&\cdots \\ 0&0&0&0&5/11&0&\cdots \\ 
 \vdots&\vdots &\vdots&\vdots&\vdots&\vdots&\ddots\\ \end{array}\right]
\label{me3}
\end{eqnarray}
In the second line of (\ref{me3}) the row labels $l$ and column labels $l'$ include the integers $0,1,2\ldots,$
$2n-1$. The elements of all but the last row of the matrix $\lvec l|\hat\mu|l')$ sum to 1.  Using (\ref{in12cc}) with (\ref{me3})  we note the identities
\begin{equation}
\lvec 0|\hat \mu=\sum_{l=0}^{2n-1}\lvec 0|\hat\mu|l)\lvec l|=\lvec 1|,
\label{in12a}
\end{equation}
and
\begin{equation}
\lvec 0|\hat \mu^2=\lvec 1|\hat \mu=\sum_{l=0}^{2n-1}\lvec 1|\hat\mu|l)\lvec l|=\frac{\lvec 0|+2\lvec 2|}{3}.
\label{in12aa}
\end{equation}

As shown in connection with (WH-106), the $2n$ different eigenvalues $\mu_i$ of (\ref{in6}) are the roots of the Legendre polynomial $P_{2n}$, that is,
\begin{equation}
P_{2n}(\mu_i)=0.\label{in6a}
\end{equation}
We will order the roots such that 
\begin{equation}
-1<\mu_1<\mu_2<\cdots<\mu_{2n}<1.
\label{in6b}
\end{equation}
We can use the eigenvalues $\mu_i$ and the eigenvectors $|\mu_i)$ and $\lvec\mu_i|$ of (\ref{in6}) to write the direction cosine matrix, $\hat\mu$ as 
\begin{equation}
\hat\mu=\sum_{i=1}^{2n}\mu_i|\mu_i)\lvec\mu_i|.
\label{in6c}
\end{equation}

The projections of the left and right multipole basis vectors $\lvec l|$ and $|l)$ onto the right and left stream bases, $|\mu_i)$ and $\lvec\mu_i|$, onto the multipole bases were give by (WH-84) and (WH-85) as
\begin{equation}
\lvec l|\mu_{i'}) =\frac{1}{2}P_l(\mu_{i'}).
\label{sdbv2}
\end{equation}
and
\begin{equation}
\lvec \mu_i|l)=w_i(2l+1)P_l(\mu_i).
\label{sdbv4}
\end{equation}
We can multiply (\ref{sdbv2}) on the right by $\lvec \mu_i|$, sum over $i$ and use (\ref{in12c}) to find the useful identity
\begin{eqnarray}
\lvec l|&=&\sum_{i=1}^{2n}\lvec l|\mu_i)\lvec \mu_i|\nonumber\\
&=&\frac{1}{2}\sum_{i=1}^{2n}P_l(\mu_i)\lvec \mu_i| 
\label{in12f}
\end{eqnarray}

The projections onto the continuous direction-cosine bases $|\mu\rangle$ and $\langle \mu|$ of (WH-26) were given by 
\begin{equation}
\langle \mu|\mu_i) = \sum_{l=0}^{2n-1}\langle \mu|l)\lvec l|\mu_{i}) =\sum_{l=0}^{2n-1}\frac{2l+1}{2}P_l(\mu)P_l(\mu_{i}).
\label{sdbv12}
\end{equation}
and
\begin{equation}
\lvec \mu_i|\mu\rangle =\sum _{l=0}^{2n-1}\lvec  \mu_i|l)\lvec l|\mu\rangle =w_i \sum_{l=0}^{2n-1}\frac{2l+1}{2} P_l(\mu_i)P_l(\mu).
\label{sdbv14}
\end{equation}
One of many equivalent formulas for the weights, $w_i$, was given by  (WH-72) as
\begin{equation}
\frac{1}{w_i}=\sum_{l=0}^{2n-1}\frac{2l+1}{2}P_l^2(\mu_i).
\label{in7a}
\end{equation}
Numerical values of the Gauss-Legendre weights, $w_i$, and cosines, $\mu$, are given in pages 916 - 919 of Abramowitz and Stegun\cite{Abramowitz}.

Eq.(\ref{in2}) also contains the efficiency matrix $\hat\eta$, defined by (WH-54) in terms of the scattering-phase matrix $\hat p$ and the single-scattering albedo $\tilde\omega$ as
\begin{equation}
\hat\eta =  \hat 1-\frac{\hat\omega}{2}\hat p.
\label{vet6}
\end{equation}
After a collision between a photon and a cloud particulate or atmospheric molecule, 
the fraction of photons scattered, as opposed to being absorbed, is the single-scattering albedo $\tilde \omega$. Since $\tilde\omega$ is a probability, it is constrained to have the values
\begin{equation}
0\le \hat\omega\le 1.
\label{vet6a}
\end{equation}
The fraction of photons absorbed and converted to heat is $1-\tilde\omega$.

Following (WH-40), we write
the scattering-phase matrix $\hat p$ of (\ref{vet6}) as
\begin{equation}
\hat p=2\sum_{l=0}^{\infty}p_l|l)\lvec l|.
\label {me12}
\end{equation}
For randomly oriented scatterers, $\hat p$ must be diagonal in multipole space, with matrix  elements $\lvec l|\hat p|l')=2p_l\delta_{ll'}$.  The coefficients $p_l$ are subject to the constraints of (WH-14) and (WH-15),
\begin{equation}
p_0=1,\quad\hbox{and}\quad |p_l|<1,\quad\hbox{for}\quad l>0.
\label {me12a}
\end{equation}
We can  use (\ref{vc14}) and (\ref{vc22}) to write  the matrix elements for axially symmetric scattering of radiation with direction cosine $\mu'$ to radiation with direction cosine $\mu$ as
\begin{eqnarray}
p(\mu,\mu') &=& \langle\mu|\hat p|\mu'\rangle\nonumber\\
&=&2\sum_{l=0}^{\infty}p_l\langle \mu|l)\lvec l|\mu'\rangle\nonumber\\
&=&\sum_{l=0}^{\infty}p_l(2l+1)P_l(\mu)P_l(\mu'),
\label {me12aa}
\end{eqnarray}
the same as (WH-11).   We define the phase function for vertical incident radiation, with $\mu'=1$ as
\begin{eqnarray}
p(\mu) &=& p(\mu, 1)\nonumber\\
&=&\sum_{l=0}^{\infty}p_l(2l+1)P_l(\mu)
\label {me12d}
\end{eqnarray}
The context should prevent any confusion between the phase function $p(\mu,\mu')$ for scattering from $\mu'$ to $\mu$ and the phase function  $p(\mu) =p(\mu,1)$ for scattering from $\mu'=1$ to $\mu$.

A simple example is the phase function for polarization-averaged Rayleigh scattering 
\begin{equation}
p(\mu)= \frac{3}{4}(1+\mu^2)= P_0(\mu)+\frac{1}{2}P_2(\mu).
\label{rs2}
\end{equation}
The non-zero coefficients of the multipole expansion (\ref{me12a}) are
\begin{equation}
p_0=1,\quad\hbox{and}\quad p_2=\frac{1}{10}.
\label{rs4}
\end{equation}
Other important examples are the maximum-forward-scattering phases, $p(\mu)=\varpi^{\{p\}}(\mu)$ of (WH-134), which can be written as 
\begin{eqnarray}
\varpi^{\{p\}}(\mu) &=&\frac{2(1+\mu)}{p(p+1)}\left[\frac{dP_p}{d\mu}(\mu)\right]^2\nonumber\\
&=&\sum_{l=0}^{2p-1}P_l(\mu)(2l+1)\varpi^{\{p\}}_{l}.
\label{pfb2}
\end{eqnarray}
Representative  multipole coefficients  of (\ref{pfb2}),
\begin{eqnarray}
\varpi^{\{p\}}_l =\frac{1}{2}\int_{-1}^1 d\mu \,P_l(\mu)\varpi^{\{p\}}(\mu),
\label{pfb2a}
\end{eqnarray}
are listed in Table 1 of WH. 

Substituting (\ref{me12}) into (\ref{vet6}) we see that the eigenvectors of $\hat\eta$ are the multipole bases 
$|l)$ and $\lvec l|$,
\begin{equation}
\hat \eta=\sum_{l=0}^{\infty}\eta_l|l)\lvec l|,\quad\hbox{with}\quad \hat\eta|l)=\eta_l|l),\quad\hbox{and}\quad \lvec l|\hat\eta =\lvec\eta|\eta_l.
\label {me12b}
\end{equation}
In view of (\ref{me12a}) and (\ref{vet6a}),  we see that the values of the efficiency eigenvalues $\eta_l$ must be nonnegative
\begin{equation}
\eta_l = 1-\tilde\omega p_l \ge 0
\label {me12c}
\end{equation}
For future reference, we note the special cases of (\ref{me12b}) and (\ref{me12c})  
\begin{equation}
\hat \eta |0)=\eta_0|0)=(1-\tilde\omega)|0) \quad\hbox{and}\quad \lvec 0|\hat \eta =\lvec 0|\eta_0 = \lvec 0|(1-\tilde \omega). \label{in12}
\end{equation}

\section{Conservation of energy}
According to (WH-19), the  volume energy density of the radiation, $u(\tau)$, at the optical depth $\tau$ is proportional to the monopole moment $I_0(\tau)$  of the intensity,
\begin{equation}
u(\tau)=\frac{2\pi}{c}\int_{-1}^{1}d\mu I(\mu,\tau)=\frac{4\pi}{c}I_0(\tau) =\lvec 0|I(\tau)\}.\label{in5a}
\end{equation}
Here the energy density vector $|u(\tau)\}$ is related to the intensity vector $|I(\tau)\}$ by
\begin{equation}
|u(\tau)\}=\frac{4\pi}{c}|I(\tau)\},\label{in5aa}
\end{equation}
where $c$ is the speed of light. 

In (WH-20) the vertical flux $Z(\tau)$ of radiation is shown to be proportional to the dipole moment $I_1(\tau)$ of the intensity
\begin{eqnarray}
Z(\tau)=2\pi\int_{-1}^{1}d\mu \,\mu I(\mu,\tau)= 4\pi I_1=\lvec 0|Z(\tau)\}.\label{in5b}
\end{eqnarray}
The flux vector $|Z(\tau)\}$ is related to the intensity vector $|I(\tau)\}$ by
\begin{equation}
|Z(\tau)\}=4\pi \hat\mu|I(\tau)\},\label{in5c}
\end{equation}
where $\hat\mu$ is the direction cosine matrix of (\ref{me3}).  Unlike the energy density $u(\tau)$ of (\ref{in5a}) which is always non-negative, $u(\tau)\ge 0$, the  flux $Z(\tau)$ can be positive for net upward radiation or negative for net downward radiation.

For conservative scattering, the second moment of the intensity is of special significance. Following \S{\bf 11} (86) of Chandrasekhar\cite{Chandrasekhar}, we write the second moment in terms of the ``$K$-integral,"
\begin{eqnarray}
K(\tau)=\frac{1}{2}\int_{-1}^{1}d\mu \,\mu^2 I(\mu,\tau)= \frac{I_0+2 I_2}{3}=\lvec 0|K(\tau)\}.\label{in5bb}
\end{eqnarray}
Here the $K$-integral vector $|K(\tau)\}$ is related to the intensity vector $|I(\tau)\}$ by
\begin{equation}
|K(\tau)\}= \hat\mu^2|I(\tau)\}.\label{in5bbb}
\end{equation}

Multiplying  (\ref{in2}) on the left by $\lvec 0|$, using (\ref{in5}) and  (\ref{in12}) with (\ref{in12a}) we find
\begin{equation}
\frac{d}{d\tau}Z(\tau)+(1-\tilde\omega)c u(\tau)=4\pi (1-\tilde\omega)B(\tau).\label{in14}
\end{equation}
For time-independent conditions, thermal radiation may heat or cool the atmosphere because of an excess of absorption over emission or vice versa. Over most of Earth's troposphere radiation cooling is approximately balanced by convective heating.  The heating rate $R(\tau)$ per unit of optical depth is defined by
\begin{equation}
R(\tau)\, d\tau = Z(\tau)-Z(\tau+d\tau) = -d\tau\frac{dZ(\tau)}{d\tau},\quad
\hbox{or}\quad R(\tau)=-\frac{dZ(\tau)}{d\tau}.\label{in14a}
\end{equation}
The heating $R(\tau)d\tau$ of the infinitesimally thin layer between the optical depths $\tau$ and $d\tau$ is the excess  of the flux $Z(\tau)$ through the bottom surface of the layer over the flux $Z(\tau+d\tau)$ through the top surface.
Using  (\ref{in14}) with (\ref{in14a}) we find that the heating rate can be written as
\begin{equation}
R (\tau)=(1-\tilde\omega)[c u(\tau)-4\pi B(\tau)].\label{in14b}
\end{equation}
The absorption rate of energy is $(1-\tilde\omega)c u(\tau)$ and the emission rate is $(1-\tilde\omega)4\pi B(\tau)$.

\begin{figure}[t]
\postscriptscale{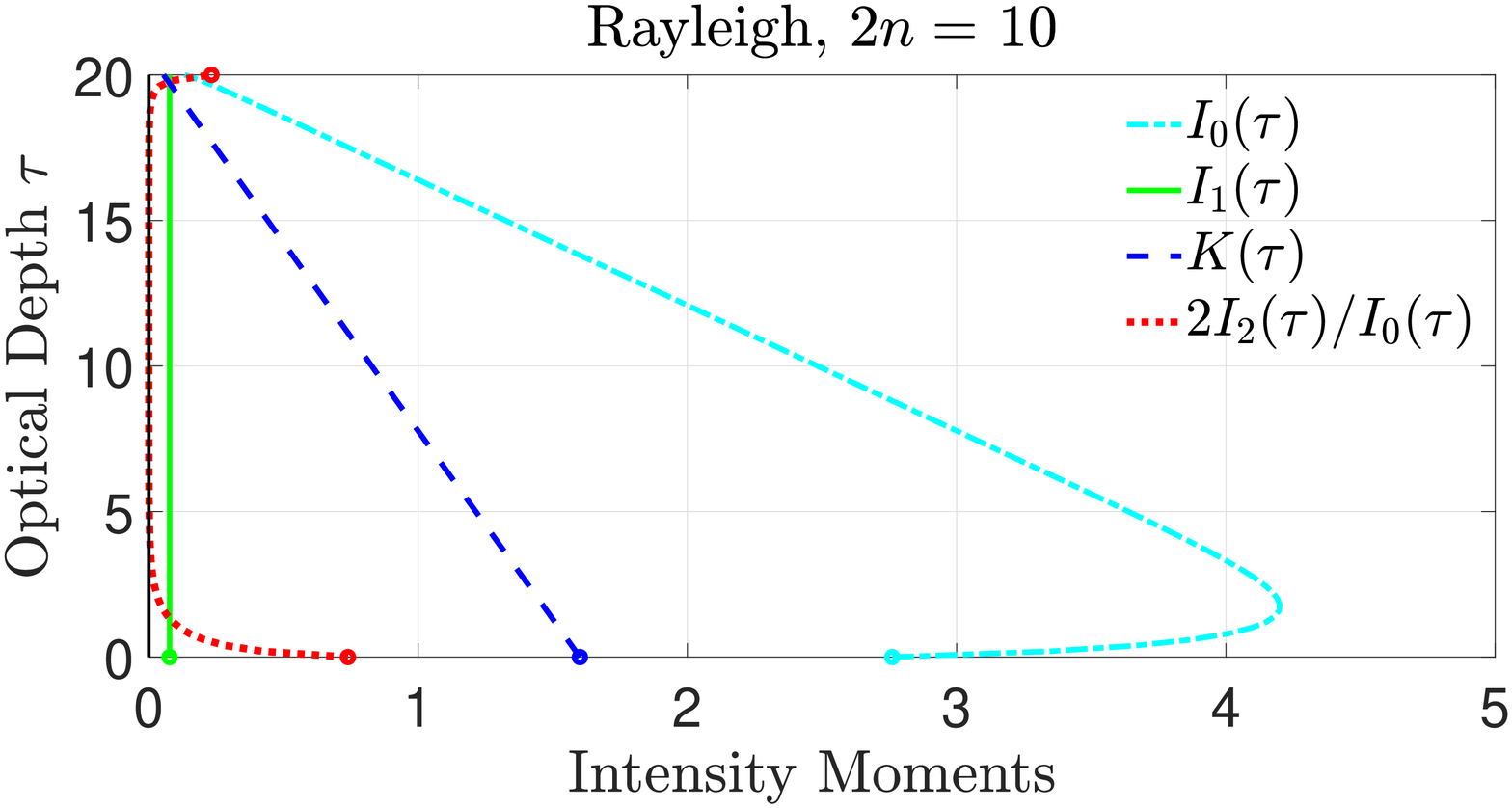}{1}
\caption{ Moments of the  intensity vector  $|I(\tau)\}$  inside  a non-absorbing, 
Rayleigh-scattering cloud of total optical depth $\tau_c = 20$, calculated with (\ref{iic2}), versus optical depth $\tau$ above the bottom of the cloud. The multipole moments are $I_l(\tau)=\lvec l|I(\tau)\}$, for $l=0,1,2$ and the value of the $K$-integral is given by (\ref{in5bb}). The input intensity vector $|I^{\rm in}\}=|\mu_{10}) 2/\mu_{10}$  has a direction cosine $\mu_{10}=0.9739$,  the maximum possible for $2n = 10$ streams. The dipole moment of the input intensity is $I_1^{\rm in}=\lvec 1|I^{\rm in}\}=1$. Because of diffuse reflection, the first moment $I_1$ of the intensity inside the cloud is much less, $I_1=0.0773$. Only 7.73\% of the input flux is difffusely transmitted through the cloud, and the remaining 92.27\% is reflected from the bottom. The dashed red line is a measure of how much the intensity violates the Eddington approximation, $I_2=0$. See the text for more details.
\label{mom2t20v}}
\end{figure}

\begin{figure}[t]
\postscriptscale{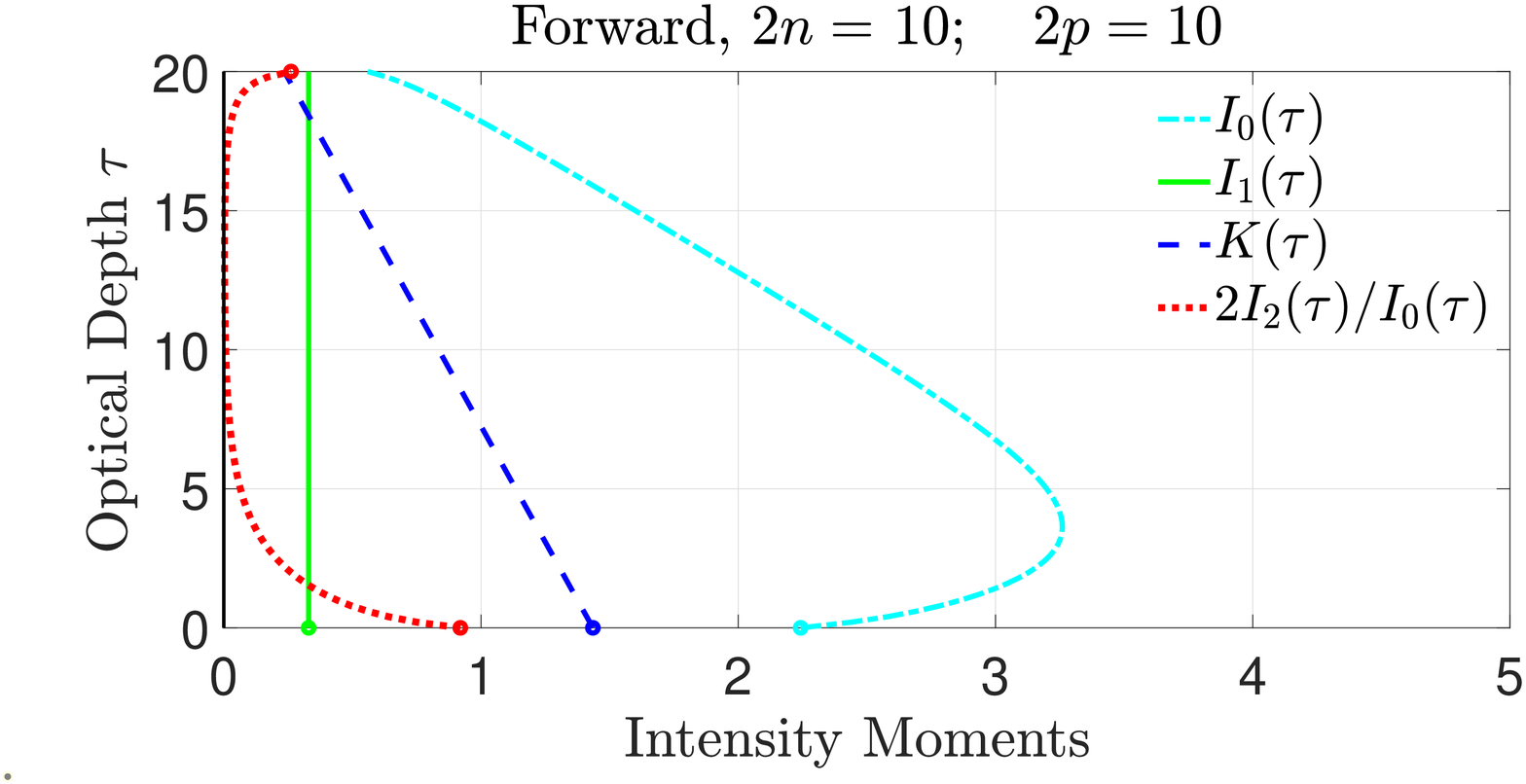}{1}
\caption{ Like Fig. \ref{mom2t20v}, but for a forward scattering phase function 
$p(\mu)=\varpi^{\{p\}}(\mu)$ of (WH-134) with $p=5$. According to (WH-138), the peak forward scattering is $p(1) =p(p+1)= 30$, compared to $p(1) = 1.5$ for Rayleigh scattering, in accordance with (WH-132).  Because the incident photons penetrate more deeply into the cloud before their first scattering, compared to the nearly isotropic Rayleigh scattering of Fig. \ref{mom2t20v}, and since the photons have a harder time reversing direction by multiple scattering,  the diffuse transmission is substantially greater than for Fig. \ref{mom2t20v}, 32.99\% compared to 7.73\%. The diffuse reflection is correspondingly less, 67.01\% compared to 92.27\%.
\label{mom3t20v}}
\end{figure}

For conservative scattering,  $\tilde\omega =1$ so (\ref{in14}) implies that $dZ(\tau)/d\tau = 0$. Then the flux $Z(\tau)$ of (\ref{in5b}), and therefore the dipole moment of the intensity, $I_1(\tau)$, are independent of $\tau$.
\begin{equation}
Z(\tau) =Z(0),\quad\hbox{and}\quad I_1(\tau)=I_1(0).
\label{in14c}
\end{equation}
Examples of the $\tau$-independence of  $I_1(\tau)$ in clouds with conservative scattering are shown as the vertical green lines of Figs. \ref{mom2t20v}, \ref{mom3t20v}, \ref{mom3t20h} and \ref{mom3t20i}.

Multiplying  (\ref{in2}) on the left by $\lvec 1|$  and using (\ref{in5}) and (\ref{me12b}) with (\ref{in5bbb}) 
we find
\begin{eqnarray}
0&=&\lvec 1|\hat\mu\frac{d}{d\tau}|I (\tau)\}+\eta_1 I_1(\tau)\nonumber\\
&=&\lvec 0|\hat\mu^2\frac{d}{d\tau}|I(\tau) \}+\eta_1 I_1(\tau) \nonumber\\
&=&\frac{dK(\tau)}{d\tau}+\eta_1 I_1(\tau).\label{in20}
\end{eqnarray}
Since $I_1(\tau)$ is independent of $\tau$, in accordance with (\ref{in14c}), we can solve (\ref{in20}) to find that the $K$-integral changes linearly with optical depth $\tau$.
\begin{equation}
K(\tau)= K(0)-\tau\eta_1 I_1(0). \label{ps16}
\end{equation}
Examples of the linear dependence of $K(\tau)$ on $\tau$ for clouds with conservative scattering are shown as the dashed blue  lines of Figs. \ref{mom2t20v}, \ref{mom3t20v}, \ref{mom3t20h} and \ref{mom3t20i}.
Chandresekhar\cite{Chandrasekhar} discusses the basic features of conservative scattering outlined above in his sections \S{\bf 8}-{\bf 10}. 
\subsection{Propagation \label{dd}}
From now on we will consider radiation in clouds that are cold enough to neglect thermal emission. Sunlight in Earth's clouds is an example of such a situation. Then we can set $B(\tau)=0$ on the right side of (\ref{in2}).
In analogy to (\ref{in6c}) we write the direction-secant matrix
\begin{eqnarray}
\hat\varsigma&=&\hat\mu^{-1}\nonumber\\
&=&\sum_{i=1}^{2n}\varsigma_i|\mu_i)\lvec\mu_i|,\quad\hbox{where}\quad
\varsigma_i =\frac{1}{\mu_i},
\label{dd0}
\end{eqnarray}
Multiplying (\ref{in2}) on the left by $\hat\varsigma$ from (\ref{dd0})
and recalling that the exponentiation rate matrix is
\begin{equation}
\hat\kappa=\hat\varsigma \hat\eta
\label{dd2a}
\end{equation}
we find
\begin{equation}
\left(\frac{d}{d\tau}+\hat\kappa\right)|I (\tau)\}=\hat 0.
\label{dd2}
\end{equation}
If we multiply (\ref{dd2}) on the left by $\lvec \mu_i|$, and assume isotropic scattering with $\hat p = 2|0)\lvec 0|$, we find the fundamental  Eq. (7) of Wick's pioneering paper\cite{Wick} on $2n$-stream scattering, which he used to analyze thermal neutron diffusion with isotropic scattering cross sections.
 
The multipole-space representations (WH-44) of $\hat\varsigma$, and (\ref{me12b}) of $\hat\eta$, allow us to write the exponentiation rate matrix of (\ref{dd2}) as
\begin{eqnarray}
\lvec l|\hat\kappa|l')&=&\lvec l|\hat\varsigma|l')\eta_{l'}\nonumber\\
&=&\left[\begin{array}{rrrrrrr} \eta_0&3\eta_1&0&-14\eta_3/3&0&88\eta_5/15&\cdots \\ 0&0&0&0&0&0&\cdots \\ 
 0&0&0&7\eta_3/3&0&-44\eta_5/15&\cdots \\   0&0&5\eta_2/3&0&0&0&\cdots\\ 0&0&0&0&0&11\eta_5/5&\cdots \\ 0&0&-4\eta_2/3&0&9\eta_4/5&0&\cdots \\ 
 \vdots&\vdots &\vdots&\vdots&\vdots&\vdots&\ddots\\ \end{array}\right].
\label{dd4}
\end{eqnarray}
From inspection of (\ref{dd4}), and for future reference, we note the identity
\begin{equation}
\lvec 0|\hat\kappa|l)+2\lvec 2|\hat\kappa|l)=\eta_0\delta_{0l}+3\eta_1\delta_{1l}.
\label{dd6}
\end{equation}
One can use Bonnet's recursion formula of (WH-18) to write a formal proof of  (\ref{dd6}).

In the simple analysis of WH, eigenvectors of the {\it propagation-length matrix}, $\hat \lambda = \hat\kappa^{-1}$, were used to describe how the intensity vector $|I(\tau)\}$ depends on $\tau$.  This procedure fails for conservative scattering because $\eta_0=1-\tilde\omega\to 0$ as $\tilde\omega\to 1$ and therefore all elements of the first column of (\ref{dd4}) vanish, $\lvec l|\hat\kappa|0)\to 0$, as $\tilde\omega\to 1$.  Then $\det \hat \kappa\to 0$  and $\hat \kappa$ cannot be inverted. Here we show how to get around this problem.

For further analysis we will assume that $\eta_0=0$ and we will find $2n$ independent solutions of (\ref{dd2}), which we denote by $|v_1), |v_2),|v_3),\ldots,|v_{2n})$. All of these $2n\times 1$ {\it propagation basis vectors}, $|v_i)=|v_i(\tau)\big)$, depend on $\tau$ and  satisfy the equation
\begin{equation}
\left(\frac{d}{d\tau}+\hat\kappa\right)|v_i)=\hat 0,
\label{dd10}
\end{equation}
Then the intensity vector $|I(\tau)\}$ can be written as
\begin{equation}
|I(\tau)\}=\sum_{i=1}^{2n}|v_i)A_i
\label{dd12}
\end{equation}
The  values of the $\tau$-independent amplitudes $A_i$ are  determined by boundary conditions.
We can represent the bases $|v_i)$ as $\tau$-dependent, $2n\times 1$  column  vectors with multipole amplitudes $\lvec l|v_i)$,
\begin{equation}
|v_i)=\left[\begin{array}{r}\lvec 0|v_i)\\ \lvec 1|v_i)\\\lvec 2|v_i) \\ \vdots\\\lvec 2n-1|v_i)\end{array}\right].
\label{dd8}
\end{equation}

One simple solution to (\ref{dd10}) which can be found by inspection of (\ref{dd4}) is
\begin{equation}
|v_1\big)=3\eta_1\tau|0)-|1)=\left[\begin{array}{r}3\eta_1\tau\\-1\\0 \\ \vdots\end{array}\right].
\label{dd14}
\end{equation}
For $|v_{1})$ the energy density $u$ of (\ref{in5a})  is
\begin{equation}
 u = \frac{4\pi\lvec 0|v_{1}\big)}{c} = \frac{12\pi\eta_1 \tau}{c}. 
\label{dd14a}
\end{equation}
The  flux, $Z(\tau)$ of (\ref{in5b}),  is 
\begin{equation}
 Z (\tau)= 4\pi\lvec 1|v_{1}\big)= -4\pi.
\label{dd14b}
\end{equation}
The $K$-integral $K(\tau)$ (\ref{in5bb}) of $|v_{2n})$ can be evaluated with (\ref {in12aa}) and is
\begin{equation}
K(\tau)=\lvec 0|\hat\mu^2|v_{1})=\frac{\lvec 0|v_{1})+2\lvec 2|v_{1})}{3}=\eta_1\tau.
\label{dd14c}
\end{equation}
Thus, $|v_{1})$ represents  a $\tau$-independent downward flux, $Z(\tau)=-4\pi$. The energy density $u(\tau)$ and the  $K$-integral  $K(\tau)$ have maximum values at the cloud top, where $\tau = \tau_c$,  and they decrease linearly to zero at the cloud bottom where $\tau = 0$. The basis vector $|v_{1})$ of (\ref{dd14}) is mentioned as 
\S {\bf 10}(80)  by Chandrasekhar\cite{Chandrasekhar}. 
 Our multipole coefficient $p_1$ is his $\varpi_1/3$, since Chandrasekhar does not include a statistical weight in his multipole expansion of the scattering phase,  \S {\bf 3}(33).

\begin{figure}[t]
\postscriptscale{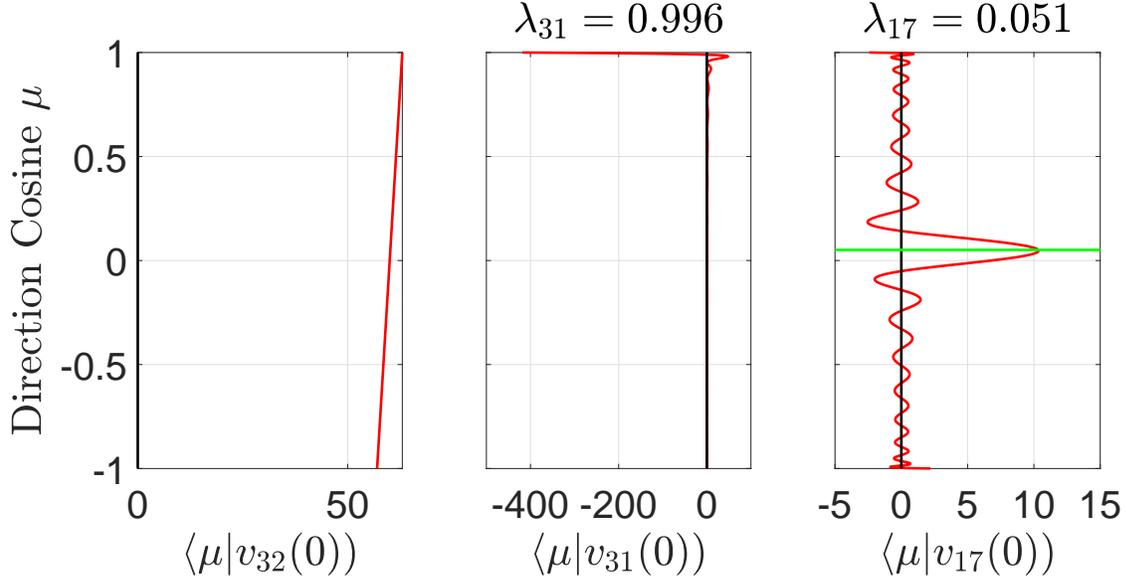}{1}
\caption{Representative projections  $\langle \mu |v_i(\tau))$ of (\ref{dd60}) at the cloud bottom where  $\tau=0$. The cloud  thickness is $\tau_c=20$. Rayleigh scattering is modeled with $2n=32$ streams. The value of the quasi-isotropic basis vector $\langle\mu|v_{32}(0)\big)$ of (\ref{dd16}), shown in the left panel, varies linearly  with the direction cosines $\mu$. The middle panel shows a nearly vertical directional basis vector $\langle\mu|v_{31}(0)\big)$ with a penetration length $\lambda_i = 0.981$. The right panel shows a nearly horizontal directional basis vector $\langle\mu|v_{17}(0)\big)$ with a penetration length $\lambda_{17} = 0.051$, indicated by the horizontal green line.  See the text for more details.
\label{muv2}}
\end{figure}

\begin{figure}[t]
\postscriptscale{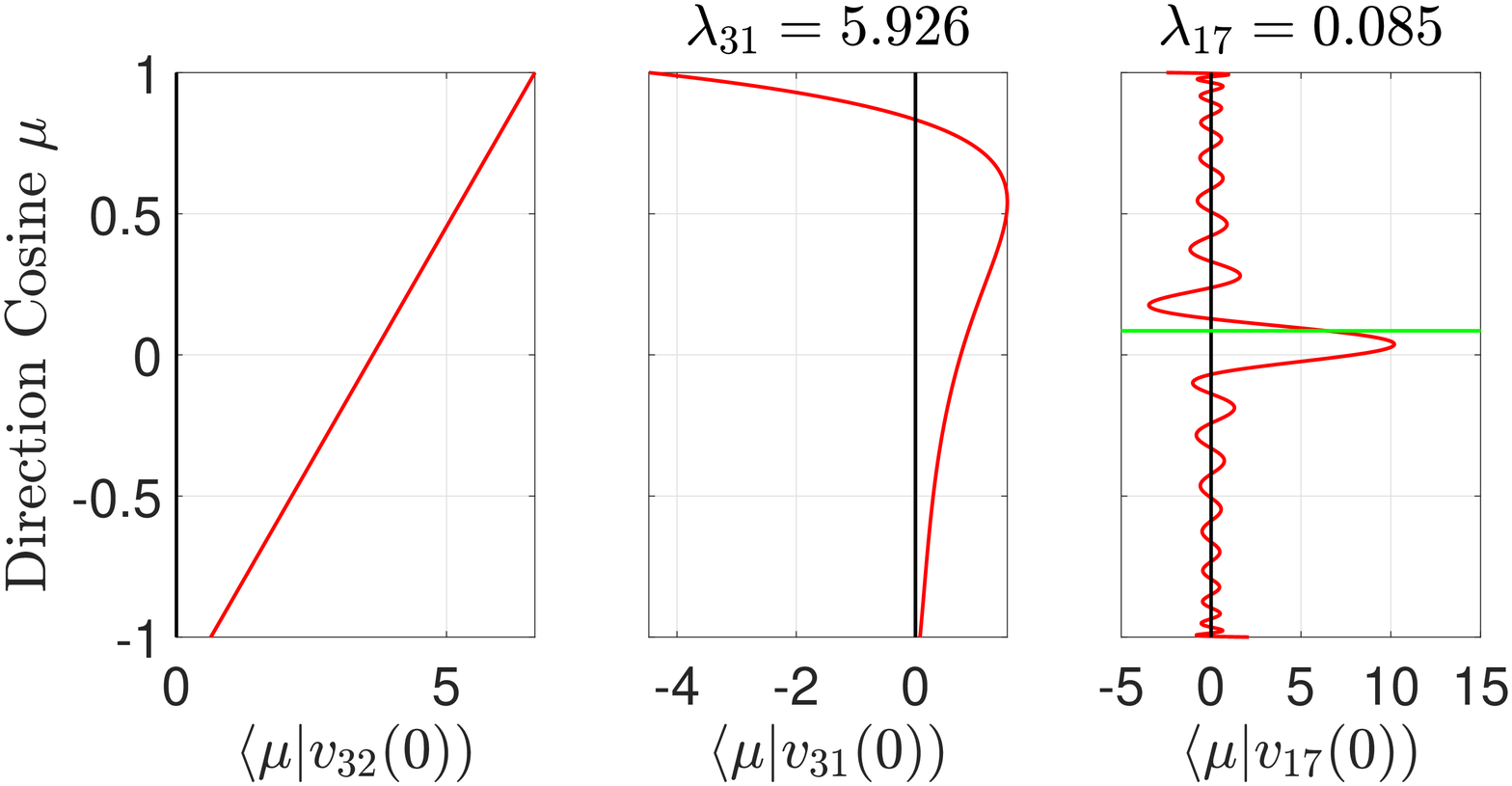}{1}
\caption{ Like Fig. \ref{muv2} but for a strongly peaked forward-scattering phase function $p(\mu)=\varpi^{\{16\}}(\mu)$, constructed from the first $2p=32$ Legendre polynomials and with a forward scattering phase of $p(1)=p(p+1)=272$.
The cloud  thickness is $\tau_c=20$, and  there are $2n =32$ streams. The value of the quasi-isotropic basis vector $\langle\mu|v_{32})$ of (\ref{dd16}), shown in the left panel, varies linearly with  $\mu$.
The middle panel shows a nearly vertical directional basis vector $\langle\mu|v_{31}(0)\big)$ with a penetration length $\lambda_i = 5.926$. The right panel shows a nearly horizontal directional basis vector $\langle\mu|v_{17}(0)\big)$ with a penetration length $\lambda_{17} = 0.085$, indicated by the horizontal green line. See the text for more details.
\label{muv3}}
\end{figure}

An analog of (\ref{dd14}), which also satisfies (\ref{dd10}), is 
\begin{equation}
|v_{2n}\big)=3\eta_1 (\tau_{c}-\tau) |0)+|1)=\left[\begin{array}{r}3\eta_1(\tau_{c}-\tau)\\1\\0 \\ \vdots\end{array}\right].
\label{dd16}
\end{equation}
For $|v_{2n})$ the energy density $u(\tau)$ of (\ref{in5a})  is
\begin{equation}
 u(\tau)= \frac{4\pi\lvec 0|v_{2n}\big)}{c} = \frac{12\pi\eta_1(\tau_c- \tau)}{c}. 
\label{dd16a}
\end{equation}
The  flux, $Z(\tau)$ of (\ref{in5b})  is 
\begin{equation}
 Z(\tau) = 4\pi\lvec 1|v_{2n}\big)= 4\pi.
\label{dd16b}
\end{equation}
The $K$-integral $K(\tau)$ (\ref{in5bb}) of $|v_{2n})$ can be evaluated with (\ref {in12aa}) and is
\begin{equation}
K(\tau)=\lvec 0|\hat\mu^2|v_{2n})=\frac{\lvec 0|v_{2n})+2\lvec 2|v_{2n})}{3}=\eta_1(\tau_c-\tau).
\label{dd14c}
\end{equation}
Thus, $|v_{2n})$ represents  a $\tau$-independent upward flux, $Z(\tau)=4\pi$. The energy density $u(\tau)$ and the  $K$-integral $K(\tau)$ have maximum values at the cloud bottom, where $\tau = 0$,  and they decrease linearly to zero at the cloud top where $\tau = \tau_c$.
Since the bases $|v_1)$ and $|v_{2n})$ contain only isotropic S waves with $l=0$ and nearly isotropic P waves with $l=1$,  we will call them the two quasi-isotropic bases, to distinguish them from the remaining  basis vectors $|v_2), |v_3),\ldots, |v_{2n-1})$, which we will call directional bases. Examples of quasi-isotropic and directional bases are shown in Fig. \ref{muv2} and \ref{muv3}. 

Numerical studies suggest that the directional bases $|v_i)$, with $i=2,3,\ldots,2n-1$, can be written as 
the sum of an isotropic part $|v_i^{I})$  and a directional part $|v_i^{D})$,
\begin{equation}
|v_i)=|v_i^{I}) + |v_i^{D}).
\label{dd18}
\end{equation}
The isotropic part is pure S-wave,
\begin{equation}
|v_i^I)=|0)\lvec 0|v_i).
\label{dd18a}
\end{equation}
The directional part is waves of multipolarity $l\ge 2$,
\begin{equation}
|v_i^{D})=\sum_{l=2}^{2n-1}|l)\lvec l|v_i)
\label{dd18b}
\end{equation}
As we will prove below,
the directional bases have no P-wave part,
\begin{equation}
\lvec 1|v_i)=0.
\label{dd18c}
\end{equation}
Therefore the directional bases $|v_i)$, for $i=2,3,\ldots, 2n-1$, carry no flux, $Z=4\pi\lvec 1|v_i) =0$.

Then (\ref{dd10}) becomes
\begin{equation}
\left(\frac{d}{d\tau}+
\left[\begin{array}{rr} \hat \kappa^{I}&\hat \kappa^{X}\\\hat 0&\hat \kappa^{D}\end{array}\right]\right)\left[\begin{array}{r}|v_i^{I}) \\|v_i^{D})\end{array}\right]=\hat 0.
\label{dd20}
\end{equation}
From inspection of (\ref{dd4}) we see that
\begin{equation}
\hat \kappa^{I}=
\left[\begin{array}{rr} 0&3\eta_1\\0&0\end{array}\right]
=3\eta_1|0)\lvec 1|.
\label{dd22}
\end{equation}
The $2\times (2n-2)$ matrix that couples directional components $|v_i^{D})$ of $|v_i)$ to quasi-isotropic components is
\begin{equation}
\hat \kappa^{X}=
\left[\begin{array}{ccccc} 0&-14\eta_3/3&0&88\eta_5/15&\cdots \\ 0&0&0&0&\cdots \end{array}\right]
=\sum_{l=2}^{2n-1}|0)\lvec 0|\hat \kappa|l)\lvec l|.
\label{dd24}
\end{equation}
The $(2n-2)\times(2n-2)$ matrix that couples directional components of $|v_i)$ to each other is
\begin{equation}
\hat \kappa^{D}=
\left[\begin{array}{rrrrr} 
 0&7\eta_3/3&0&-44\eta_5/15&\cdots \\   5\eta_2/3&0&0&0&\cdots\\ 0&0&0&11\eta_5/5&\cdots \\ -4\eta_2/3&0&9\eta_4/5&0&\cdots \\ 
 \vdots&\vdots&\vdots&\vdots&\ddots\\ \end{array}\right]
=\sum_{l\ge 2,l'\ge 2}|l)\lvec l|\kappa|l')\lvec l'|.
\label{dd26}
\end{equation}
The lowest row of the block-matrix equation (\ref{dd20}) can be written as the $(2n-2)\times(2n-2)$ matrix equation,
\begin{equation}
\left(\frac{d}{d\tau}+\kappa^{D}\right)|v_i^{D})=\hat 0.
\label{dd28}
\end{equation}
The matrix $\kappa^{D}$ will have
 $(2n-2)$ right eigenvectors $|\lambda_i)$ and non-zero eigenvalues 
 $\kappa_i$, defined, aside from normalization by
\begin{equation}
\hat\kappa^D|\lambda_i)=\kappa_i|\lambda_i).
\label{dd30}
\end{equation}
It will be convenient to label the eigenvectors $|\lambda_i)$ with the inverses $\lambda_i$ of the eigenvalues $\kappa_i$,
\begin{equation}
\lambda_i = \frac{1}{\kappa_i}
\label{dd30a}
\end{equation}
We will call the inverse eigenvalues {\it penetration lengths}, and we will assume that they are ordered such that
\begin{equation}
\lambda_2<\lambda_3<\cdots<\lambda_{2n-1}.
\label{dd32}
\end{equation}
Since $\kappa^{D}$ is odd under reflection, $\hat r \kappa^{D}\hat r = -\kappa^{D}$, 
where the reflection operator $\hat r$ of (WH-116) can be written as
\begin{equation}
\hat r =\sum_{l=0}^{2n-1}(-1)^l|l)\lvec l|,
\label{dd17a}
\end{equation}
the eigenvectors can be chosen to be reflection conjugates of each other
\begin{equation}
\hat r|\lambda_i)=|\lambda_{r(i)}),\quad\hbox{where}\quad r(i)=2n-i+1.
\label{dd38}
\end{equation}
The corresponding eigenvalues must be
 equal and opposite,
\begin{equation}
\lambda_i=-\lambda_{r(i)}.
\label{dd39}
\end{equation}

We guess that the isotropic and directional parts of the 
bases $|v_i)$ are
\begin{equation}
\left[\begin{array}{r}|v_i^{I}) \\|v_i^{D})\end{array}\right]
=\left[\begin{array}{c}-2|0)\lvec 2|\lambda_i)  \\   |\lambda_i)\end{array}\right]e^{-\kappa_i(\tau-\tau_i)},
\label{dd40}
\end{equation}
or
\begin{equation}
\lvec l|v_{i})
=\left[\begin{array}{r}-2\lvec 2|\lambda_i)  \\ 0\\ \lvec 2  |\lambda_i)\\ \lvec 3 |\lambda_i)\\ \vdots\\ 
\lvec 2n-1  |\lambda_i)\end{array}\right]
e^{-\kappa_i(\tau-\tau_i)}.
\label{dd41}
\end{equation}
The amplitude of the isotropic, S-wave part of $|v_i)$ is  $\lvec 0|v_i) = -2\lvec 2|\lambda_i)$.

In (\ref{dd40}) and (\ref{dd41}) the reference optical depths $\tau_i$ are
\begin{equation}
\tau_i=\left \{\begin{array}{rl}\tau_c, &\mbox{if $i=2,3,4,\ldots,n$, }\\
0,&\mbox{if $i = n+1,n+2,n+3,\ldots, 2n-1$.} \end{array}\right . 
\label{dd42}
\end{equation}
To prove that the bases $|v_i)$ of (\ref{dd41}) are solutions of (\ref{dd10})
we first note that differentiating (\ref{dd41}) gives
\begin{equation}
\frac{d}{d\tau}|v_{i})=-\kappa_i|v_i).
\label{dd43}
\end{equation}
Noting from (\ref{dd22}) and (\ref{dd41}) that $\hat \kappa^I|v_i^I)=0$ we see that
\begin{equation}
\hat\kappa |v_{i})=\left[\begin{array}{rr} \hat \kappa^{I}&\hat \kappa^{X}\\\hat 0&\hat \kappa^{D}\end{array}\right]\left[\begin{array}{c}|v_{i}^I) \\|v_{i}^{D})\end{array}\right]
=\left[\begin{array}{c}\hat\kappa^X|v_{i}^{D}) \\ \hat\kappa^D|v_{i}^{D})\end{array}\right].
\label{dd44}
\end{equation}
The bottom element on the right of (\ref{dd44}) is
\begin{equation}
\hat\kappa^D|v_{i}^{D})=\hat\kappa^D|\lambda_i)e^{-\kappa_i(\tau-\tau_i)}
=\kappa_i|\lambda_i)e^{-\kappa_i(\tau-\tau_i)}=\kappa_i|v_{i}^{D}).
\label{dd46}
\end{equation}
The top element on the right of (\ref{dd44}) is
\begin{eqnarray}
\hat\kappa^X|v_{i}^{D})&=&\sum_{l=2}^{2n-1}|0)\lvec 0|\hat\kappa|l)\lvec l|\lambda_i)e^{-\kappa_i(\tau-\tau_i)}\nonumber\\
&=&-2\sum_{l=2}^{2n-1}|0)\lvec 2|\hat\kappa^D|l)\lvec l|\lambda_i)e^{-\kappa_i(\tau-\tau_i)}\nonumber\\
&=&-2|0)\lvec 2|\hat\kappa^D|\lambda_i)e^{-\kappa_i(\tau-\tau_i)}\nonumber\\
&=&-2\kappa_i|0)\lvec 2|\lambda_i)e^{-\kappa_i(\tau-\tau_i)}\nonumber\\
&=&\kappa_i|v^I_i).
\label{dd48}
\end{eqnarray}
The top line of (\ref{dd48}) comes from (\ref{dd24}). To get the second line we noted that for $l\ge 2$
we can use (\ref{dd6}) to write $\lvec 0|\hat\kappa|l)=-2\lvec 2|\hat \kappa|l)=-2\lvec 2|\hat \kappa^D|l)$. The sum on $|l)\lvec l|$ then becomes a unit operator in directional space, which leads to the third line. The fourth line follows from (\ref{dd30}). The final line follows from (\ref{dd40}).
Using (\ref{dd48}) and (\ref{dd46}) in (\ref{dd44}) we find
\begin{equation}
\hat\kappa |v_{i})=\kappa_i|v_i).
\label{dd50}
\end{equation}
Adding (\ref{dd50}) to (\ref{dd43}) we find (\ref{dd10}).  

The eigenvalue equations (\ref{dd30}) do not determine how the eigenvectors $|\lambda_i)$ are normalized.  It will be convenient to let the top element have the value 1/2,
\begin{equation}
-2\lvec 2 |\lambda_i)=\frac{1}{2}.
\label{dd51}
\end{equation}
Then we can write the directional basis vectors $|v_i)$ of (\ref{dd41}), with $i=2,3,\ldots, 2n-1$, as
\begin{equation}
|v_{i})=\left[\begin{array}{r}\lvec 0 |v_i) \\ \lvec 1 |v_i)\\  \lvec 2 |v_i)\\ \lvec 3 |v_i)\\ \vdots\\ 
\lvec 2n-1  |v_i)\end{array}\right]
=\left[\begin{array}{r}0.5 \\ 0\\ -0.25\\ \lvec 3 |\lambda_i)\\ \vdots\\ 
\lvec 2n-1  |\lambda_i)\end{array}\right]
e^{-\kappa_i(\tau-\tau_i)}.
\label{dd52}
\end{equation}

For the directional basis vectors  $|v_{i})$ of (\ref{dd52}), the energy density $u(\tau)$, given by (\ref{in5a}),  decays exponentially with distance into the cloud from the reference optical depth $\tau_i$,
\begin{equation}
u = \frac{4\pi}{c} \lvec 0|v_{i})= \frac{2\pi}{c}e^{-\kappa_i(\tau-\tau_i)}.
\label{dd52a}
\end{equation}
The  flux, $Z(\tau)$, given by (\ref{in5b}), is 
\begin{equation}
Z=4\pi\lvec 1|v_{i}) = 0. 
\label{dd52b}
\end{equation}
The  $K$-integral $K(\tau)$ given by (\ref{in5bb}) is also zero for the directional bases,
\begin{equation}
K(\tau)=\frac{\lvec 0|v_{2n})+2\lvec 0|v_{2n})}{3}=0.
\label{dd52c}
\end{equation}

Calculations of radiation transfer in stars sometimes use the  {\it Eddington approximation},
\begin{equation}
\lvec l|I(\tau)\} =I_l(\tau)= 0,\quad\hbox{for}\quad l\ge 2.
\label{dd53}
\end{equation}
An example of how valid the Eddington criterion (\ref{dd53}) is for a realistic cloud can be seen in Fig. \ref{mom2t20v}. For optical depths from about $\tau =2$ above the bottom of the cloud to about 2 optical depths below the top of the cloud at $\tau=20$, the Eddington approximation is very well satisfied, and $2I_2(\tau)/I_0(\tau)\approx 0$. But the figure shows that the Eddington approximation is not good near the bottom of the cloud, where the highly directional input radiation is being isotropized by multiple scattering, or near the top of the cloud where isotropic radiation from inside the cloud is being transformed into purely upward radiation from the top surface. Near the bottom and top of the cloud the intensity $|I(\tau)\}$ contains large fractions of directional bases (\ref{dd52}), which do not satisfy the Eddington criterion (\ref{dd53}), since they are normalized such that $\lvec 2|v_i)=-1/4$. But the directional bases decay exponentially to zero with increasing distances from the top and bottom of the cloud. Eddington's criterion is very well satisfied near the center of an optically thick cloud, where the radiation is mostly due to the quasi-isotropic bases $|v_{2n})$ and $|v_1)$,  of (\ref{dd14}) and (\ref{dd16}), which identically satisfy the Eddington criterion, $I_l(\tau) = 0$ for $l\ge 2$.

Fig. \ref{mom2t20v} also shows the monopole moment, $I_0(\tau)$, or average intensity. Isotropization of near-normal-incidence radiation causes the average intensity to maximize  just above the bottom of the cloud. Qualitatively similar increases of nearly isotropic neutron thermal fluxes are observed because of slowing down of fast neutrons in efficient neutron reflectors. \cite{neutrons}

The simplest model with both quasi-isotropic and directional basis vectors has $n=2$ stream pairs. Then  the basis vectors are
\begin{equation}
\lvec l |v_i)=\left[\begin{array}{r r r r}3\eta_1\tau&0.5\,e^{-\kappa_2(\tau-\tau_c)}&0.5\,e^{-\kappa_3\tau}
&3\eta_1(\tau_c-\tau)\\ -1&0&0&1\\0&-0.25\,e^{-\kappa_2(\tau-\tau_c)}&-0.25\, e^{-\kappa_3\tau}&0\\0&\lvec 3|\lambda_2)\, e^{-\kappa_2(\tau-\tau_c)}&\lvec 3|\lambda_3)\, e^{-\kappa_3\tau}&0\end{array}\right]
\label{dd54}
\end{equation}
In accordance with the symmetry (\ref{dd39}),
\begin{equation}
-\kappa_2=\kappa_{3}=\frac{\sqrt{35\eta_2\eta_3}}{3}.
\label{dd56}
\end{equation}
In accordance with the symmetry (\ref{dd38}),
\begin{equation}
\lvec 3 |\lambda_2)=-\lvec 3|\lambda_{3})=\frac{1}{4}\sqrt{\frac{5\eta_2}{7\eta_3}}.
\label{dd58}
\end{equation}

Figure (\ref{muv2}) shows representative projections  
\begin{eqnarray}
\langle\mu|v_i(\tau)\big) &=& \sum_{l=0}^{2n-1}\langle \mu|l)\lvec l|v_i(\tau)\big)\nonumber\\
&=& \sum_{l=0}^{2n-1}P_l(\mu)(2l+1)\lvec l|v_i(\tau)\big).
\label{dd60}
\end{eqnarray}
of the basis vectors $|v_i(\tau)\big)$ of (\ref{dd8}) onto the continuous $\mu$-space basis $\langle\mu|$, with $\langle\mu|l)$ given by (\ref{vc22}). As can be seen in Fig. \ref{muv2}, for $|\lambda_i|\le 1$, the directional bases $\langle\mu|v_i)$ have their largest values  for $\mu\approx \lambda_i$.  A basis vector with large amplitudes for $\mu\approx 1$ represents nearly vertical radiation which penetrates the cloud better than radiation from a basis vector representing nearly horizontal radiation, for which the largest values of $\langle\mu|v_i)$ occur for $\mu\approx 0$. Horizontal radiation must traverse a longer slant path through the cloud than vertical radiation, and is attenuated more.  So basis vectors $|v_i)$ representing the  most horizontal radiation have the shortest penetration lengths $\lambda_i$.

\begin{figure}[t]
\postscriptscale{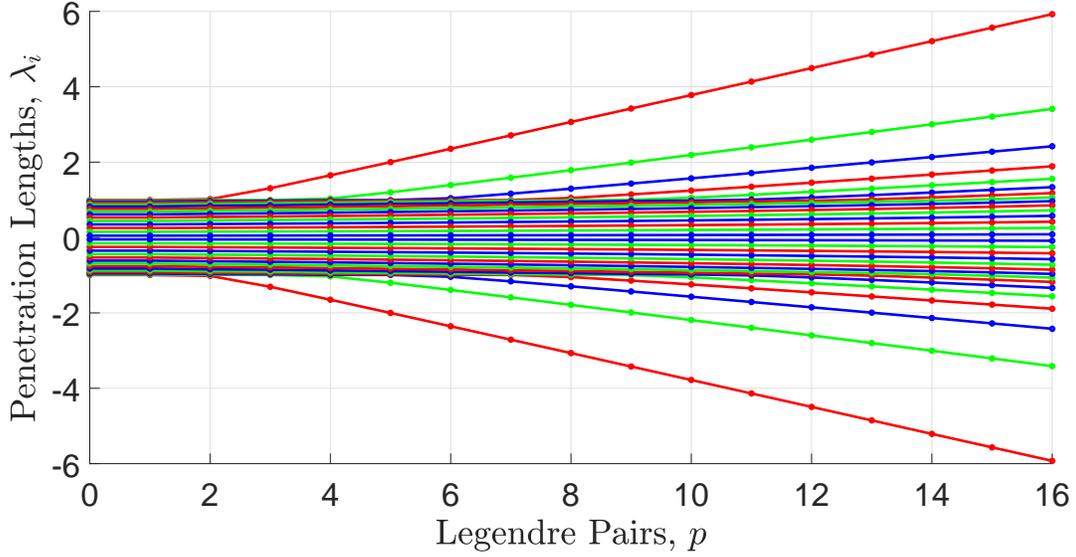}{1}
\caption{The $2n-2=30$ penetration lengths $\lambda_i$ of (\ref{dd30a}) for a conservative scattering model with $2n=32$ streams, and with the phase functions $p(\mu)=\varpi^{\{p\}}(\mu)$ of (\ref{pfb2}). These phase functions give the maximum forward scattering possible for functions constructed from the first $2p$ Legendre polynomials. The possible values of $p$ are $1,2,3,\ldots, 16$. The values of $\lambda_i$ labeled with $p=0$ are for purely isotropic scattering, with $p(\mu) = 1$.
\label{lamip}}
\end{figure}
For  phase functions $\varpi^{\{p\}}(\mu)$ of (\ref{pfb2}) for $p\gg 1$ and strong forward scattering, the largest penetration lengths $|\lambda_i|$ of (\ref{pfb2}) can exceed 1, as can be seen in Fig. \ref{lamip}. The directional basis vectors $|v_i)$, corresponding to $\lambda_i>1$, are somewhat delocalized, but are largest near $\mu =1$. The basis vectors corresponding to $\lambda_i<-1$ are largest near $\mu = -1$. Figure \ref{muv3} shows representative projections for strongly forward-peaked phase functions.  
\section{The scattering  matrix}
For a cloud with a bottom at $\tau=0$ and a top at $\tau=\tau_c$, the incoming and outgoing intensity vectors are
\begin{equation}
|I^{\{\rm in\}} \} =\mathcal{M}_{\bf u}|I(0)\}+\mathcal{M}_{\bf d}|I(\tau_c)\}, 
\label{c10}
\end{equation}
and
\begin{equation}
|I^{\{\rm out\}} \} =
\mathcal{M}_{\bf d}|I(0)\} +\mathcal{M}_{\bf u} |I(\tau_c)\}.
\label{c12}
\end{equation}
as discussed in (WH-174) and (WH-175).
The projection matrices $\mathcal{M}_{\bf d}$ and $\mathcal{M}_{\bf u}$ for downward streams with indices $j\le n$ and upward streams with indices $k>n$, respectively, are given by (WH-88) as
\begin{equation}
\mathcal{M}_{\bf d }=\sum_{j=1}^{n}|\mu_j)\lvec \mu_j|,\quad\hbox{and}\quad
\mathcal{M}_{\bf u} =\sum_{k=n+1}^{2n}|\mu_k)\lvec \mu_k|.
\label{sdbv9}
\end{equation}
The projection matrices have the simple algebra
\begin{eqnarray}
\mathcal{M}_{\bf d }+\mathcal{M}_{\bf u} &=&\hat 1,\label{sdbv9a}\\
\mathcal{M}^2_{\bf d }=\mathcal{M}_{\bf d}\quad&\hbox{and}&\quad
\mathcal{M}^2_{\bf u}=\mathcal{M}_{\bf u} ,\label{sdbv9b}\\
\mathcal{M}_{\bf d }\mathcal{M}_{\bf u}=\hat 0\quad&\hbox{and}&\quad
\mathcal{M}_{\bf u}\mathcal{M}_{\bf d} =\hat 0,\label{sdbv9c}
\end{eqnarray}
For future reference, we use (\ref{sdbv9a}) -- (\ref{sdbv9c}) to show that the  inverses of (\ref{c10}) and (\ref{c12}) are
\begin{equation}
|I(0)\} =\mathcal{M}_{\bf u} |I^{\{\rm in\}}\}+
\mathcal{M}_{\bf d}|I^{\{\rm out\}} \} .
\label{c14a}
\end{equation}
\begin{equation}
|I(\tau_c)\} =\mathcal{M}_{\bf d} |I^{\{\rm in\}}\}+
\mathcal{M}_{\bf u}|I^{\{\rm out\}} \} .
\label{c14b}
\end{equation}

Multiplying (\ref{c12}) on the left by $\lvec\mu_i|$, and using (\ref{dd12}) we see that the intensity of the $i$th outgoing stream is
\begin{eqnarray}
\lvec\mu_i|I^{\{\rm out\}} \} 
&=&\sum_{i'}\left[\lvec\mu_i|\mathcal{M}_{\bf d}|v_{i'}(0)\big)+\lvec\mu_i|\mathcal{M}_{\bf u}|v_{i'}(\tau_c)\big)\right]A_{i'}\nonumber\\
&=&\sum_{i'}\mathcal{O}_{ii'}A_{i'}
\label{sam2}
\end{eqnarray}
The elements of the outgoing  matrix , with $j\le n$ and $k>n$,  are
\begin{eqnarray}
\mathcal{O}_{ji'}&=&\lvec\mu_j|v_{i'}(0)\big)=\sum_l\lvec\mu_j|l)\lvec l|v_{i'}(0)\big)\nonumber\\
\mathcal{O}_{ki'}&=&\lvec\mu_k|v_{i'}(\tau_c)\big)=\sum_l\lvec\mu_k|l)\lvec l|v_{i'}(\tau_c)\big)\
\label{sam4}
\end{eqnarray}
\begin{figure}[t]
\postscriptscale{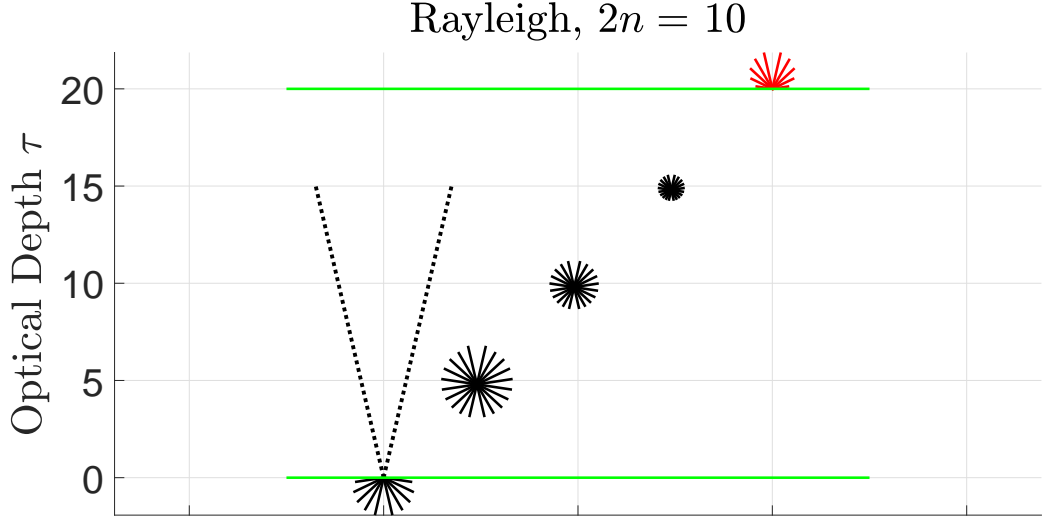}{1}
\caption{Intensities inside the Rayleigh-scattering cloud of Fig. \ref{mom2t20v}. The ray lengths are proportional to the unweighted intensities $I(\mu_i,\tau)=w_i^{-1}\lvec \mu_i|I(\tau)\}$, calculated with (\ref{iic2}), at the optical depth $\tau$. At the bottom of the cloud, with $\tau =0$, there is an upward incident stream $|I^{\{\rm in\}}=|\mu_{10})2/\mu_{10}$, indicated by dotted rays, and diffusely reflected downward streams, indicated by the continuous rays.  Within a few optical depths above the bottom, the intensity isotropizes due to multiple scattering, and there is little trace of the directional incident stream.  The red color of the outgoing rays at the top of the cloud, with $\tau = \tau_c = 20$, indicates that they have been magnified by a factor of 10 for clarity.
Only upward streams exist from the top of the cloud since we assumed only upward incident light.  If the transmission at the top of the cloud or the reflection at the bottom were Lambertian, the rays would have the same lengths. But in both cases there is substantial ``limb darkening.'' The nearly vertical rays are stronger than the nearly horizontal rays.}\label{ctr2t20v}
\end{figure}
\begin{figure}[t]
\postscriptscale{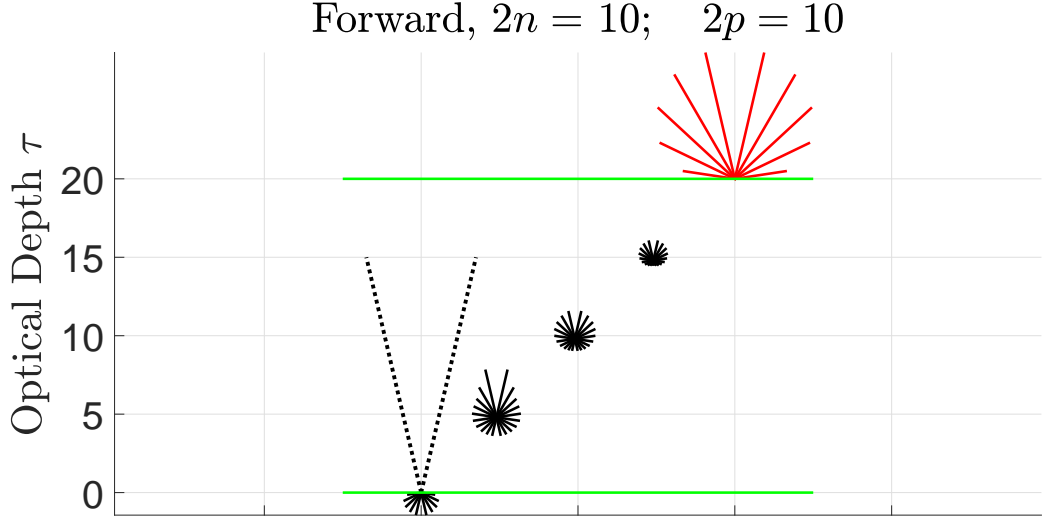}{1}
\caption{Unweighted intensities $I(\mu_i,\tau)=w_i^{-1}\lvec \mu_i|I(\tau)\}$ like those of  Fig. \ref{ctr2t20v} but for a forward scattering phase function 
$p(\mu)=\varpi^{\{p\}}(\mu)$ of (WH-134) with $p=5$. According to (WH-138), the peak forward scattering is $p(1) =p(p+1)= 30$.  Because the incident photons penetrate more deeply into the cloud before their first scattering, compared to the nearly isotropic Rayleigh scattering of Fig. \ref{ctr2t20v}, and because the photons have a harder time reversing direction by multiple scattering, the diffuse reflection is substantially less, and the diffuse transmission is substantially greater than for Fig. \ref{ctr2t20v}, as discussed in the caption of Fig. \ref{mom3t20v}. Even at an optical depth $\tau =5$ above the bottom of the cloud, the intensity has not fully isotropized.  Both the diffuse transmission and reflection are strongly limb darkened.
}\label{ctr3t20v}
\end{figure}
\begin{figure}[t]
\postscriptscale{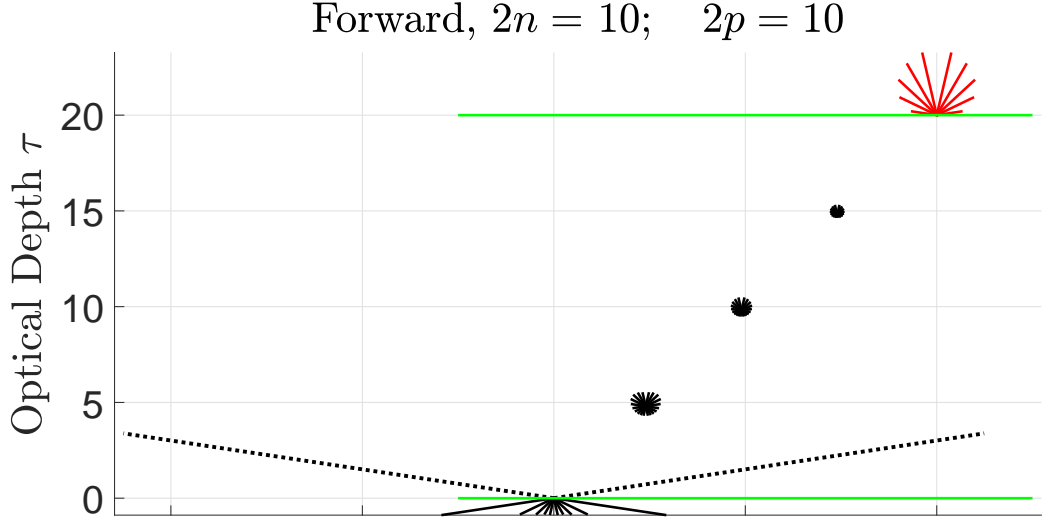}{1}
\caption{Unweighted intensities $I(\mu_i,\tau)=w_i^{-1}\lvec \mu_i|I(\tau)\}$ like 
those of Fig. \ref{ctr3t20v} but with the most horizontal incident radiation possible for $2n = 10$ streams. At the bottom of the cloud, with $\tau =0$, there is a nearly-horizontal incident stream $|I^{\{\rm in\}}\}=|\mu_{6})2/\mu_{6}$, indicated with the dotted rays, and diffusely reflected downward streams. The direction cosine of the input stream is $\mu_6 = 0.1489$. The diffusely transmitted light at the top of the cloud is limb darkened, with more intense vertical  than horizontal radiation. But at the bottom, the diffuse reflection is strongly limb brightened since the incident light is so nearly horizontal, and there is strong forward scattering. The transmission of the cloud, 13.41\%, is much less than for a nearly vertical incident beam like that of Fig. \ref{ctr3t20v} or Fig. \ref{mom3t20v},  where the transmission  is 32.99\%.
}\label{ctr3t20h}
\end{figure}
\begin{figure}[t]
\postscriptscale{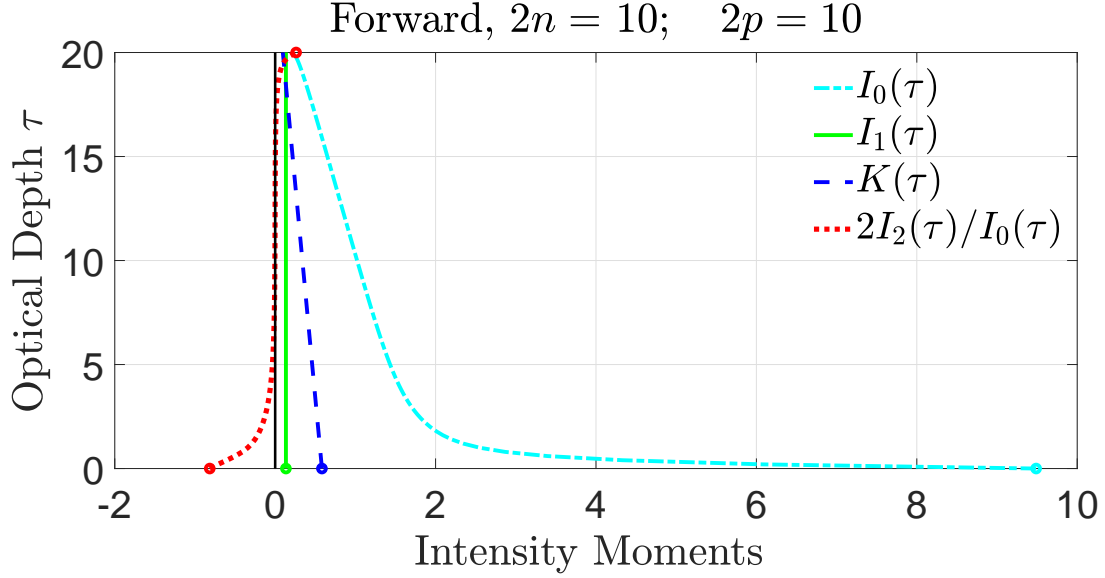}{1}
\caption{ Like Fig. \ref{mom3t20v} but  with the most horizontal incident radiation possible for $2n = 10$ streams, as in Fig.  \ref{ctr3t20h}. The input intensity vector is $|I^{\{\rm in\}}\}=|\mu_{6})2/\mu_{6}$,  the direction cosine of the input stream is $\mu_6 = 0.1489$, and the input dipole moment of the intensity vector is
 $I^{\{\rm in\}}_1=\lvec 1|I^{\{\rm in\}}\}=1$.  The energy density, $u=4\pi I_0/c$, peaks at the bottom surface of the cloud, rather than several optical depths inside, as for vertical incident radiation.  The diffuse transmission has been reduced from the 32.99\% of Fig. \ref{mom3t20v} to 13.41\%.
\label{mom3t20h}}
\end{figure}
\begin{figure}[t]
\postscriptscale{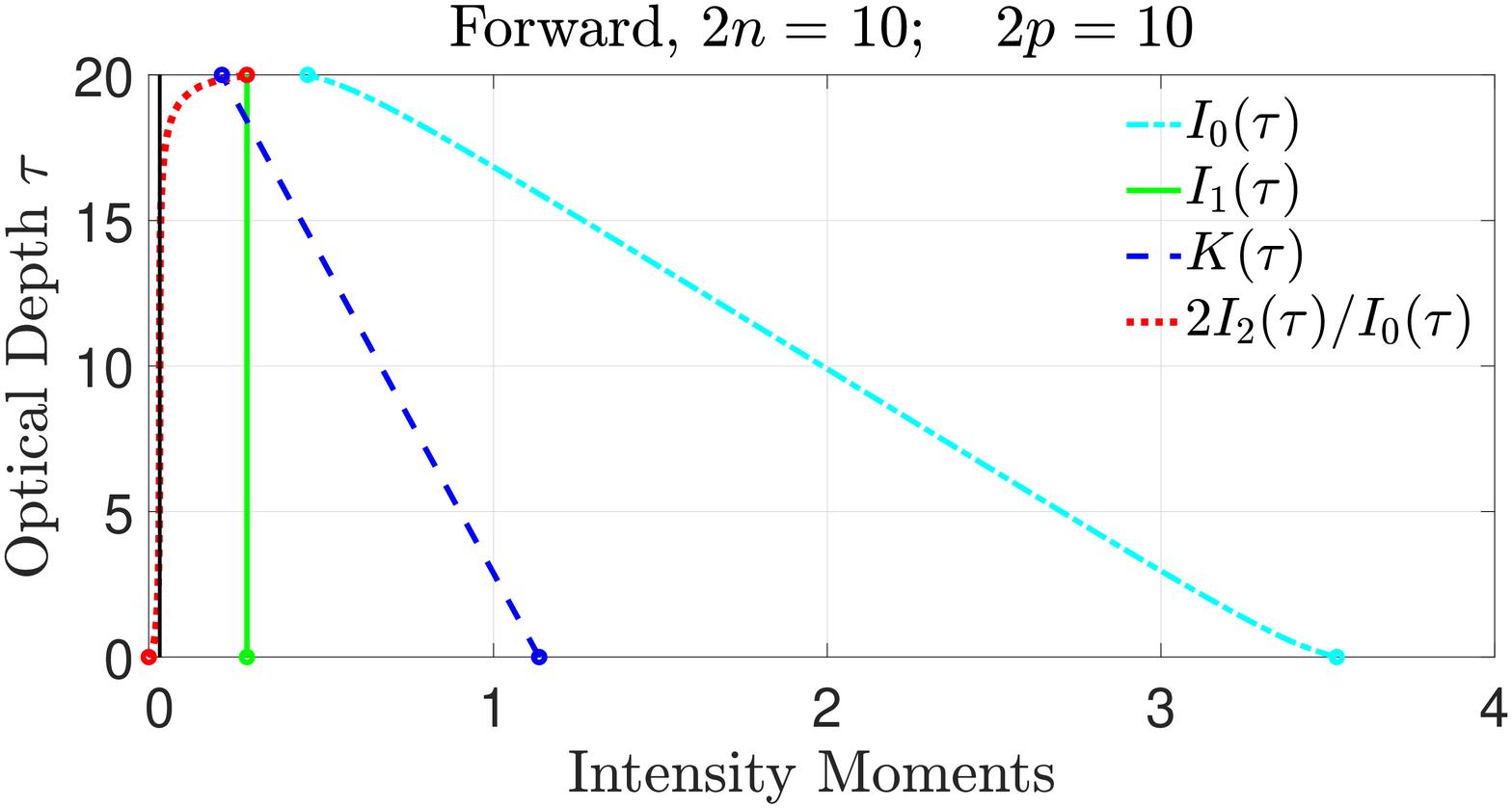}{1}
\caption{Intensity moments inside a cloud with a forward scattering phase function $p(\mu)=\varpi^{\{p\}}(\mu)$ of (\ref{pfb2}),  constructed from the first $2p=10$ Legendre polynomials.  There is half-isotropic input radiation at the bottom, $|I^{\rm in}\}=\mathcal{M}_{\bf u}|0)/\lvec 1|\mathcal{M}_{\bf u}|0)$, normalized such that 
$I_1^{\{\rm in\}}=\lvec 1|I^{\rm in}\}=1$.  The transmission, $I_1(\tau)$, is 26.14\%, intermediate between the transmission, 32.99\% of the most vertical possible stream of Fig. \ref{mom3t20v}, and the transmission 13.41\% of the most horizontal possible beam of Fig. \ref{mom3t20h}. The radiation stream directions inside the cloud are shown by the red rays of Fig. \ref{isin20}.
\label{mom3t20i}}
\end{figure}
\begin{figure}[t]
\postscriptscale{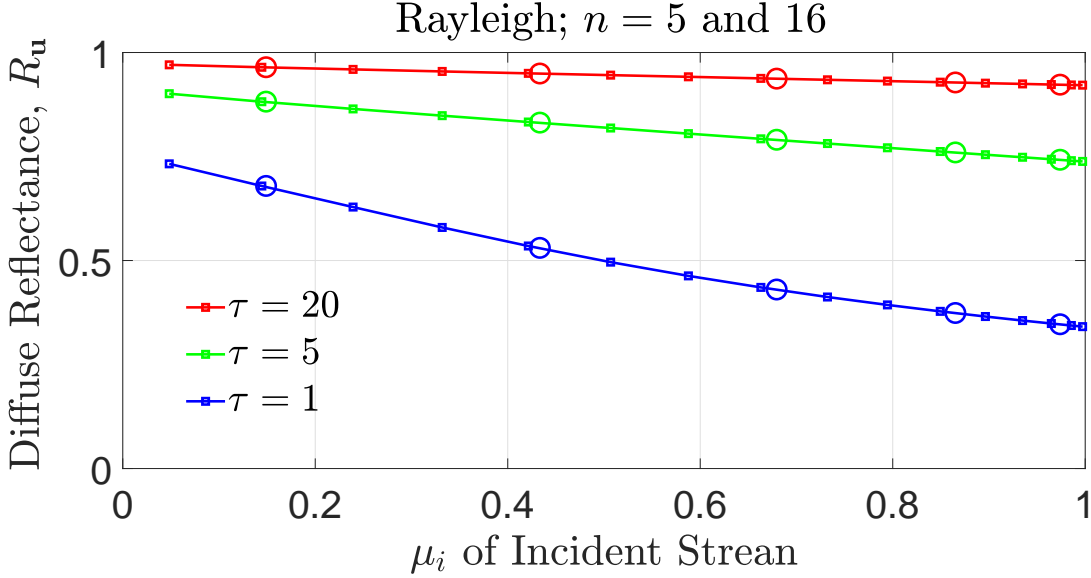}{1}
\caption{The diffuse reflectances $R_{\bf u}$ of (\ref{alb10f}) for upward incident streams, $|I^{\{\rm in\}}\}=|\mu_{i})$, with $i=n+1,n+2,\ldots, 2n$, incident on the bottom of the cloud.  Results are shown for three cloud thicknesses,
$\tau_c = 20, 5$ and $1$. The Rayleigh-scattering phase function $p(\mu)$ of (\ref{rs2}) is assumed. Results for  $2n =10$ streams (large circles) and $2n = 32$ streams (small black squares) hardly differ. The reflectance increases as the incident beams become more horizontal and as the cloud thickness increases. For conservative scattering the transmittance is $T_{\bf u}=1-R_{\bf u}$, in accordance with (\ref{alb10k}).
\label{Ref2}}
\end{figure}
\begin{figure}[t]
\postscriptscale{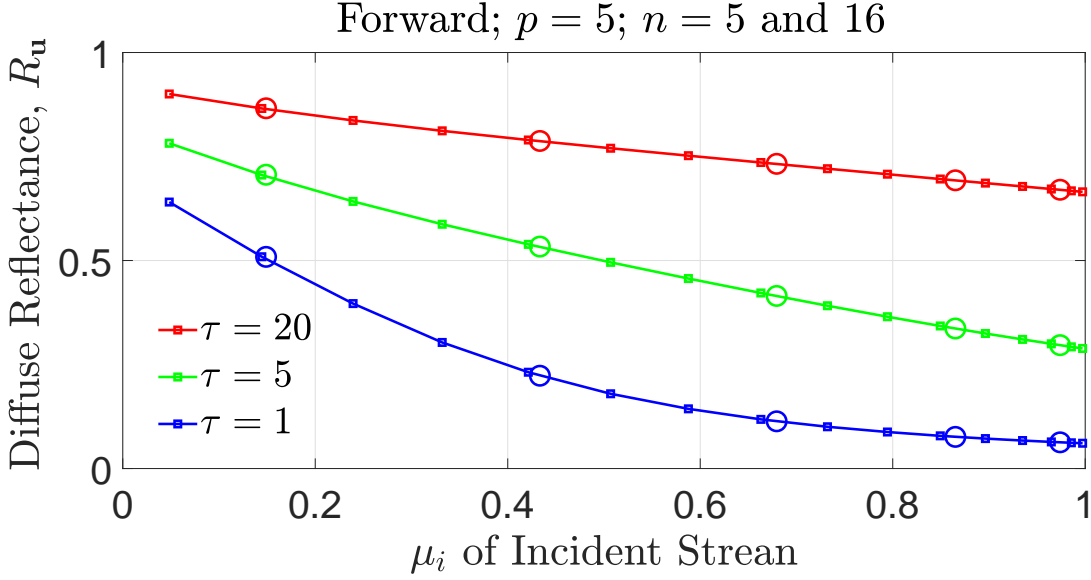}{1}
\caption{Like Fig. \ref{Ref2} but for a forward-scattering phase function $p(\mu)=\varpi^{\{p\}}(\mu)$ of (\ref{pfb2}), with $p=5$. The forward-scattering amplitude,  $p(1) = p(p+1) = 30$, is twenty times larger than that for Rayleigh scattering, $p(1)=1.5$.  Because of the forward scattering, the stream reflectances $R_{\bf u}$ are less and transmissivities, $T_{\bf u}=1-R_{\bf u}$, are greater than for Fig. \ref{Ref2}.
\label{Ref3}}
\end{figure}
\begin{figure}[t]
\postscriptscale{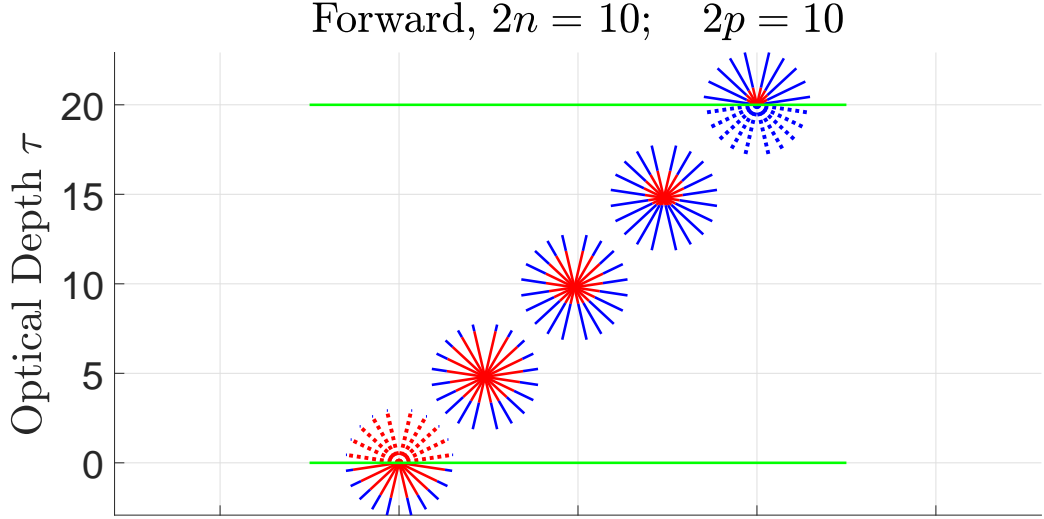}{1}
\caption{A cloud with the maximum forward scattering phase function $p(\mu)=\varpi^{\{p\}}(\mu)$ of (\ref{pfb2}) constructed from $p=5$ pairs of Legendre polynomials. The incoming intensity vector is isotropic, $|I^{\{\rm in\}}\}=|0)$.  The optical thickness  is $\tau_c = 20$ and the radiation is modeled with $2n = 10$ streams.  Unweighted intensities $I^{\{\bf u\}}(\mu_i,\tau)=w_i^{-1}\lvec\mu_i|I^{\{\bf u\}}(\tau)\}$, originating as upward streams onto the bottom of the cloud are shown as red rays and are given by (\ref{iir14}). Unweighted intensities  
$I^{\{\bf d\}}(\mu_i,\tau)=w_i^{-1}\lvec \mu_i|I^{\{\bf d\}}(\tau)\}$
originating as downward streams onto the top of the cloud 
are shown as blue rays  and are given by (\ref{iir16}).  The top and bottom input rays are dotted. 73.86\% of the radiation incident on the bottom is diffusely reflected with 26.24\% reaching the top, and vice versa for radiation incident on the top. The moments of the red rays are shown in Fig. \ref{mom3t20i}. The total radiation remains isotropic and of equal intensity everywhere inside and outside the cloud, as discussed in the text.
\label{isin20}}
\end{figure}
\begin{figure}[t]
\postscriptscale{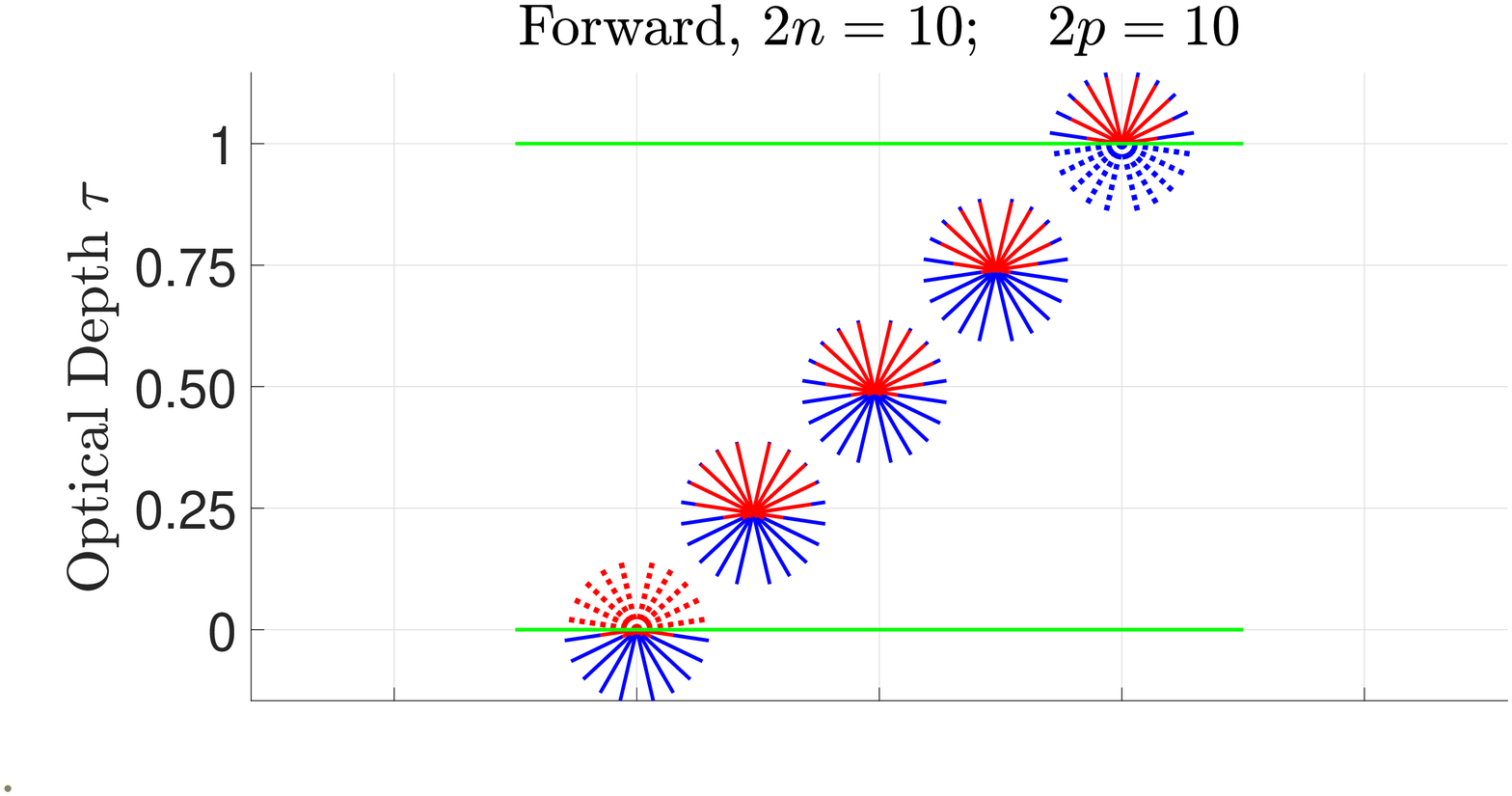}{1}
\caption{Like Fig. \ref{isin20} but for a thinner cloud, with optical thickness $\tau_c = 1$.  A smaller fraction, 15.73\%, of the ``red" radiation incident onto the bottom of the cloud is reflected, with a larger fraction, 84.27\%, transmitted at the top, and vice versa for the ``blue" radiation incident onto the top. The more vertical streams are less attenuated than the  more horizontal streams, which leads to substantial limb darkening of the red emergent radiation at the top, and the blue emergent radiation at the bottom.
\label{isin1}}
\end{figure}
Multiplying (\ref{c10}) on the left by $\lvec\mu_i|$, we see that the intensity of the $i$th incoming stream is
\begin{eqnarray}
\lvec\mu_i|I^{\{\rm in\}} \} 
&=&\sum_{i'}\left[\lvec\mu_i|\mathcal{M}_{\bf u}|v_{i'}(0)\big)+\lvec\mu_i|\mathcal{M}_{\bf d}|v_{i'}(\tau_c)\big)\right]A_{i'}\nonumber\\
&=&\sum_{i'}\mathcal{I}_{ii'}A_{i'}
\label{sam6}
\end{eqnarray}
The elements of the incoming  matrix, with $j\le n$ and $k>n$,  are
\begin{eqnarray}
\mathcal{I}_{ji'}&=&\lvec\mu_j|v_{i'}(\tau_c)\big)=\sum_l\lvec\mu_j|l)\lvec l|v_{i'}(\tau_c)\big)\nonumber\\
\mathcal{I}_{ki'}&=&\lvec\mu_k|v_{i'}(0)\big)=\sum_l\lvec\mu_k|l)\lvec l|v_{i'}(0)\big)\
\label{sam8}
\end{eqnarray}
Inverting (\ref{sam6}) we find
\begin{equation}
A_{i'}=\sum_{i''}\mathcal{I}^{-1}_{i'i''}\lvec\mu_{i''}|I^{\{\rm in\}} \} 
\label{sam10}
\end{equation}
Substituting (\ref{sam10}) into (\ref{sam2}) we find
\begin{equation}
\lvec\mu_i|I^{\{\rm out\}} \}=\sum_{i''}\mathcal{S}_{ii''}\lvec\mu_{i''}|I^{\{\rm in\}} \},\quad
\hbox{or}\quad  |I^{\{\rm out\}} \}=\mathcal{S}|I^{\{\rm in\}} \}.
\label{sam12}
\end{equation}
The elements of the scattering matrix are
\begin{equation}
\mathcal{S}_{ii''}=\sum_{i'}\mathcal{O}_{ii'}\mathcal{I}^{-1}_{i'i''},\quad\hbox{or}\quad \mathcal{S}=
\mathcal{O}\mathcal{I}^{-1}.
\label{sam12c}
\end{equation}

For radiation transfer with finite absorption, $\tilde\omega <1$, which was considered in WH, there is no need to introduce fundamentally different formulas for the quasi-isotropic bases $|v_1)$ and $|v_{2n})$, which both have a linear dependence on optical depth, and the directional bases, $|v_2)$, $|v_3),\ldots |v_{2n-1})$, which evolve exponentially with $\tau$.  With finite absorption, even the quasi-isotropic modes $|\lambda_1)$ and $|\lambda_{2n})$ evolve exponentially with $\tau$, albeit more and more slowly as $\tilde \omega\to 1$. As shown by (WH-173), the left bases, $\lvec \lambda_1|$ and $\lvec\lambda_{2n}|$ diverge as $(1-\tilde \omega)^{-1/2}$ for the conservative scattering limit, $\tilde\omega \to 1$.
\subsection{Intracloud intensities}
Using (\ref{sam10}) with (\ref{dd12}), we write the intensity vector for optical depths $\tau$ inside a cloud as
\begin{eqnarray}
|I(\tau)\}&=&\sum_{i}^{2n}|v_{i'}(\tau)\big)A_{i'}\nonumber\\
&
=&\sum_{i'i''}|v_{i'}(\tau)\big)\mathcal{I}^{-1}_{i'i''}\lvec\mu_{i''}|I^{\{\rm in\}} \} .
\label{iic2}
\end{eqnarray}
In (\ref{iic2}) and (\ref{iic4})
	the summation indices $i'$ and $i''$ take on the values $1,2,3,\ldots 2n$. 
The projections of (\ref{iic2}) onto the stream bases are
\begin{eqnarray}
\lvec\mu_i|I(\tau)\}=\sum_{l=0}^{2n-1}\lvec \mu_i|l)\lvec l|I(\tau)\}
\label{iic4}
\end{eqnarray}
Examples of (\ref{iic4}) are shown in Figs. \ref{ctr2t20v},  \ref{ctr3t20v} and  \ref{ctr3t20h} where the lengths of the rays are equal to the unweighted values 
$w_i^{-1}\lvec\mu_i|I(\tau)\}$ of the stream intensities. These correspond to the intensity moments of  Figs. \ref{mom2t20v},
\ref{mom3t20v} and  \ref{mom3t20h}.
\subsection{Identities for $\mathcal{S}$ and $\Omega$ }
The cloud albedo matrix $\Omega$, defined in (WH-216) is related to the scattering matrix  by the similarity transformation
\begin{equation}
\Omega = (\hat\mu_{\bf u}-\hat\mu_{\bf d})\mathcal{S}(\hat\varsigma_{\bf u}-\hat\varsigma_{\bf d}).
\label{c22a}
\end{equation}
The upward and downward parts of the direction-cosine matrix (\ref{in6c}) are 
\begin{equation}
\hat\mu_{\bf u}=\hat \mu\mathcal{M}_{\bf u}=\mathcal{M}_{\bf u}\hat \mu,\quad \hbox{and}\quad
\hat\mu_{\bf d}=\hat\mu\mathcal{M}_{\bf d}=\mathcal{M}_{\bf d}\hat\mu,
\label{c20a}
\end{equation}
and the upward and downward parts of the direction-secant matrix (\ref{dd0}) are 
\begin{equation}
\hat\varsigma_{\bf u}=\hat \varsigma\mathcal{M}_{\bf u}=\mathcal{M}_{\bf u}\hat \varsigma,\quad \hbox{and}\quad
\hat\varsigma_{\bf d}=\hat\varsigma\mathcal{M}_{\bf d}=\mathcal{M}_{\bf d}\hat\varsigma,
\label{c20aa}
\end{equation}
According to (\ref{in14c}), the dipole moment of the intensity, $I_1(\tau)$ is independent of the optical depth $\tau$ above the bottom of a cloud, so we can write
\begin{eqnarray}
0&=&I_1(\tau_c)-I_1(0)\nonumber\\
&=&\lvec 1|\left(\mathcal{M}_{\bf d} |I^{\{\rm in\}}\}+
\mathcal{M}_{\bf u}|I^{\{\rm out\}} \} -\mathcal{M}_{\bf u} |I^{\{\rm in\}}\}-
\mathcal{M}_{\bf d}|I^{\{\rm out\}} \} \right)\nonumber\\
&=&\lvec 0|\hat\mu\left( \mathcal{M}_{\bf d}- \mathcal{M}_{\bf u}+[\mathcal{M}_{\bf u}-
\mathcal{M}_{\bf d}]\mathcal{S}\right)|I^{\{\rm in\}} \} \nonumber\\
&=&\lvec 0|\left( \hat\mu_{\bf d}-\hat\mu_{\bf u}+[\hat\mu_{\bf u}-
\hat\mu_{\bf d}]\mathcal{S}\right)|I^{\{\rm in\}} \}.
\label{c20}
\end{eqnarray}
In the second line of (\ref{c20}) we used (\ref{in5}), (\ref{c14a}) and (\ref{c14b}). The third line follows from (\ref{in12a}) and (\ref{sam12}).  
Eq. (\ref{c20}) is true for any $|I^{\{\rm in\}}\}$ so it implies that
\begin{equation}
\lvec 0|(\hat\mu_{\bf u}-\hat\mu_{\bf d})\mathcal{S}= \lvec 0|(\hat\mu_{\bf u}-\hat\mu_{\bf d}).
\label{c22}
\end{equation}
Multiplying  both sides of (\ref{c22}) on the right by $(\hat\varsigma_{\bf u}-\hat\varsigma_{\bf d})$ and
using (\ref{c22a}) we  find,
\begin{equation}
\lvec 0|\Omega=\lvec 0|
\label{c24}
\end{equation}
The identity (\ref{c24}) can serve as  a useful consistency check for numerical calculations.

In analogy to (\ref{c24}), for conservative scattering the $S$-matrix satisfies the identity
\begin{equation}
S|0)=|0),
\label{c24a}
\end{equation}
which can also serve as a useful check of numerical calculations.
A simple proof of (\ref{c24a}) follows from Kirchhoff's law of (WH-279)
\begin{equation}
\mathcal {E} =\hat 1 -\mathcal{S}
\label{c24b}
\end{equation}
Here $\mathcal{E}$ is the isothermal emissivity matrix of the cloud.
Kirchoff's law remains valid as we approach the conservative scattering limit $\tilde \omega \to 1$, when the thermal emission of the cloud must approach zero,
\begin{equation}
\mathcal {E}|B\}=\mathcal{E}|0)B=(\hat 1-\mathcal{S})|0)B \to 0,\quad \hbox{as}\quad \tilde\omega \to 1.
\label{c24c}
\end{equation}
Here $B$ is the Planck intensity of an isothermal cloud,  given by (WH-7).
The limit  (\ref{c24c}) proves (\ref{c24a}).

\section{Transmission, reflection and absorption}
For greater clarity, we will consider transmission and reflection  by a cloud with a single scattering albedo $\tilde\omega<1$ and therefore some absorption. In the conservative scattering limit of $\tilde\omega\to 1$ the absorption vanishes.

As shown by (WH-212) and (WH-214) the incoming and outgoing flux vectors $|Z^{\{\rm in\}} \}$  and $|Z^{\{\rm out\}} \}$ are related to the incoming and outgoing intensities of (\ref{c10}) and (\ref{c12}) by
\begin{eqnarray}
|Z^{\{\rm in\}} \}&=&4\pi(\hat\mu_{\bf u}-\hat\mu_{\bf d})|I^{\{\rm in\}}\},\label{alb0g}\\
|Z^{\{\rm out\}} \}&=&4\pi(\hat\mu_{\bf u}-\hat\mu_{\bf d})|I^{\{\rm out\}}\}.
\label{alb0gg}
\end{eqnarray}
The albedo matrix $\Omega$ gives the proportionality of 
$|Z^{\{\rm out\}} \}$ to $|Z^{\{\rm in\}} \}$,
\begin{equation}
|Z^{\{\rm out\}} \}=\Omega |Z^{\{\rm in\}}\}.
\label{alb10}
\end{equation}
We will define the absorption  probability of the cloud as
\begin{equation}
A=1-\langle \Omega \rangle 
\label{alb10a}
\end{equation}
As shown in (WH-217) the expectation value $\langle \Omega\rangle$ of the cloud albedo depends on the input flux vector $|Z^{\{\rm in\}}\}$, and is given by
\begin{eqnarray}
\langle \Omega\rangle&=&\frac{Z^{\{\rm out\}}}{Z^{\{\rm in\}}}=\frac{\lvec 0|\Omega |Z^{\{\rm in\}}\}}{\lvec 0|Z^{\{\rm in\}}\}}.
\label{alb14}
\end{eqnarray}
Substituting (\ref{alb14}) into (\ref{alb10a}) we find that the absorption probability can be written as
\begin{equation}
A=A_{\bf d}f_{\bf d}+A_{\bf u}f_{\bf u}.
\label{alb10b}
\end{equation}
The fractions of downward and upward input flux are
\begin{equation}
f_{\bf d}=\frac{\lvec 0|\mathcal{M}_{\bf d}|Z^{\{\rm in\}}\}}{\lvec 0|Z^{\{\rm in\}}\}},\quad\hbox{and}
\quad f_{\bf u}=\frac{\lvec 0|\mathcal{M}_{\bf u}|Z^{\{\rm in\}}\}}{\lvec 0|Z^{\{\rm in\}}\}}.
\label{alb10c}
\end{equation}
The absorption probabilities for the downward and upward fractions of the input flux are
\begin{equation}
A_{\bf d}=\frac{\lvec 0|(1-\Omega)\mathcal{M}_{\bf d}|Z^{\{\rm in\}}\}}{\lvec 0|\mathcal{M}_{\bf d}|Z^{\{\rm in\}}\}}=1-\frac{\lvec 0|\Omega\mathcal{M}_{\bf d}|Z^{\{\rm in\}}\}}{\lvec 0|\mathcal{M}_{\bf d}|Z^{\{\rm in\}}\}}
\label{alb10d}
\end{equation}
and
\begin{equation}
A_{\bf u}=\frac{\lvec 0|(1-\Omega)\mathcal{M}_{\bf u}|Z^{\{\rm in\}}\}}{\lvec 0|\mathcal{M}_{\bf u}|Z^{\{\rm in\}}\}}=1-\frac{\lvec 0|\Omega\mathcal{M}_{\bf u}|Z^{\{\rm in\}}\}}{\lvec 0|\mathcal{M}_{\bf u}|Z^{\{\rm in\}}\}}
\label{alb10dd}
\end{equation}
From inspection of (\ref{c24}) we see that the absorption probabilities of (\ref{alb10d}) and (\ref{alb10dd}) vanish in the conservative scattering limit
\begin{equation}
A_{\bf d}\to 0,\quad\hbox{and}\quad
A_{\bf u}\to 0,\quad\hbox{as}\quad \tilde\omega \to 1.
\label{alb10e}
\end{equation}

We define the transmissivity $T_{\bf u}$ and reflectivity $R_{\bf u}$ of the upward input flux as
\begin{equation}
T_{\bf u}=\frac{\lvec 0|\mathcal{M}_{\bf u}\Omega\mathcal{M}_{\bf u}|Z^{\{\rm in\}}\}}{\lvec 0|\mathcal{M}_{\bf u}|Z^{\{\rm in\}}\}},\quad\hbox{and}\quad
R_{\bf u}=\frac{\lvec 0|\mathcal{M}_{\bf d}\Omega\mathcal{M}_{\bf u}|Z^{\{\rm in\}}\}}{\lvec 0|\mathcal{M}_{\bf u}|Z^{\{\rm in\}}\}}.
\label{alb10f}
\end{equation}
In like manner, the transmissivity $T_{\bf d}$ and reflectivity $R_{\bf d}$ of the downward input flux are
\begin{equation}
T_{\bf d}=\frac{\lvec 0|\mathcal{M}_{\bf d}\Omega\mathcal{M}_{\bf d}|Z^{\{\rm in\}}\}}{\lvec 0|\mathcal{M}_{\bf d}|Z^{\{\rm in\}}\}},\quad\hbox{and}\quad
R_{\bf d}=\frac{\lvec 0|\mathcal{M}_{\bf u}\Omega\mathcal{M}_{\bf d}|Z^{\{\rm in\}}\}}{\lvec 0|\mathcal{M}_{\bf d}|Z^{\{\rm in\}}\}}.
\label{alb10g}
\end{equation}
Summing the transmission and reflection probabilities of (\ref{alb10f}) and using (\ref{sdbv9a})  with (\ref{alb10d}) we find
\begin{eqnarray}
T_{\bf u}+R_{\bf u}&=&\frac{\lvec 0|(\mathcal{M}_{\bf u}+\mathcal{M}_{\bf d})\Omega\mathcal{M}_{\bf u}|Z^{\{\rm in\}}\}}{\lvec 0|\mathcal{M}_{\bf u}|Z^{\{\rm in\}}\}}\nonumber\\
&=&\frac{\lvec 0|\Omega\mathcal{M}_{\bf u}|Z^{\{\rm in\}}\}}{\lvec 0|\mathcal{M}_{\bf u}|Z^{\{\rm in\}}\}}\nonumber\\
&=&1-A_{\bf u},
\label{alb10h}
\end{eqnarray}
or
\begin{equation}
T_{\bf u}+R_{\bf u}+A_{\bf u}=1.
\label{alb10i}
\end{equation}
The probabilities $T_{\bf u}$ for transmission, $R_{\bf u}$ for reflection, and $A_{\bf u}$ for absorption of upward incoming flux sum to 1. In like manner, one can show that for downward incoming flux
\begin{equation}
T_{\bf d}+R_{\bf d}+A_{\bf d}=1.
\label{alb10j}
\end{equation}
For conservative scattering, when absorption probabilities $A_{\bf d}$ and $A_{\bf u}$ vanish, in accordance with (\ref{alb10e}), we can write (\ref{alb10i}) and (\ref{alb10j}) as
\begin{equation}
T_{\bf d}+R_{\bf d}=1\quad\hbox{and}\quad T_{\bf u}+R_{\bf u}=1.
\label{alb10k}
\end{equation}
\subsection{Examples}
The transmission and reflection of clouds depends on  the cloud thickness $\tau_c$, on the scattering phase function $p(\mu)$  of (\ref{me12d}), and on  the directions of the incoming radiation. The directions are described by the input intensity vector $|I^{\{\rm in\}}\}$ of (\ref{c10}) or by its equivalent, the input flux vector 
$|Z^{\{\rm in\}}\}=4\pi(\hat\mu_{\bf u}-\hat\mu_{\bf d})|I^{\{\rm in\}}\}$  of (\ref{alb0g}).

For a model with $2n=10$ streams,
Fig. \ref{mom2t20v} and Fig. \ref{ctr2t20v} show details of how the radiation penetrates a cloud of optical depth $\tau_c=20$ for the most vertical possible input stream, with $|I^{\{\rm in\}}\}=|\mu_{2n})$ and for Rayleigh scattering.

In like manner,
Fig. \ref{mom3t20v} and Fig. \ref{ctr3t20v} show details of how the radiation penetrates a cloud of optical depth $\tau_c=20$ for the most vertical possible input stream, with $|I^{\{\rm in\}}\}=|\mu_{2n})$, and for the maximum possible forward scattering, with a phase function, $p(\mu) =\varpi^{\{n\}}(\mu)$ for radiation described with $2n=10$ streams. 

Fig. \ref{ctr3t20h} and Fig. \ref{mom3t20h} show what happens if the input stream is the most horizontal upward stream possible, with $|I^{\{\rm in\}}\}=|\mu_{n+1})$.
Input streams that are nearly normal to  the surfaces of the cloud, have significantly larger diffuse transmission and smaller diffuse reflection than input streams that are more nearly horizontal, and therefore must traverse longer slant distances to get from one side of the cloud to the other.

Fig. \ref{mom3t20i} shows the moments of a half isotropic input stream $|I^{\{in\}}\} = \mathcal{M}_{\bf u}|0)$ incident on the bottom of a cloud with the same optical depth, $\tau_c =20$ as those of Fig.  \ref{ctr3t20v} for a near vertical stream and Fig.  \ref{ctr3t20h} for a near horizontal stream. The reflectivities and transmissivities are intermediate between those of near vertical and near horizontal streams. The intensity directions for Fig. \ref{mom3t20i} are shown as the red rays of Fig. \ref{isin20}. The  Eddington criterion (\ref{dd53}) is much better satisfied near the bottom of the cloud, because of the near isotropy of the intensity there.

Single-stream reflectivities calculated with (\ref{alb10f}) for upward incoming streams $|I^{\{\rm in\}}\}=|\mu_{i})$, with $i=n+1,n+2,\ldots, 2n$, for various cloud thicknesses are shown in
Fig. \ref{Ref2} for Rayleigh scattering and in Fig. \ref{Ref3} for forward scattering.
\section{Isotropic incident radiation}
An interesting special case of conservative scattering is that of isotropic input radiation, when  upward radiation is incident with equal intensities from all directions onto the bottom of the cloud, and downward radiation is incident with equal intensities onto the top of the cloud. Then we can write the incoming intensity vector
(\ref{c10}) as
\begin{equation}
|I^{\{\rm in\}} \} =|0).
\label{iir2}
\end{equation}
According to (\ref{sam12}) and (\ref{c24a}), the outgoing intensity vector is
\begin{eqnarray}
|I^{\{\rm out\}} \}&=&\mathcal{S}|I^{\{\rm in\}} \}\nonumber\\
& =&\mathcal{S}|0)\nonumber\\
&=&|0).
\label{iir4}
\end{eqnarray}
For conservative scattering of an isotropic input intensity vector, the output intensity vector is also isotropic.
From (\ref{c14a}) and (\ref{sdbv9a}) we see that the intensity vector at the bottom of the cloud is
\begin{eqnarray}
|I(0)\} &=&\mathcal{M}_{\bf u} |I^{\{\rm in\}}\}+
\mathcal{M}_{\bf d}|I^{\{\rm out\}} \} \noindent\\
&=&(\mathcal{M}_{\bf u}+
\mathcal{M}_{\bf d})|0) \noindent\\
&=&|0).
\label{iir6}
\end{eqnarray}
In like manner, we find from (\ref{c14b}) and  (\ref{sdbv9a}) that the intensity vector at the top of the cloud is
\begin{equation}
|I(\tau_c)\} = |0).
\label{iir8}
\end{equation}
Inside the cloud, with $0\le \tau\le \tau_c$ we assume that the general solution to the intensity vector, the expansion (\ref{dd12}), includes only the two quasi-isotropic bases $|v_1)$ and $|v_{2n})$. We choose the amplitudes to be
\begin{equation}
A_i=\frac{1}{3\eta_1\tau_c}(\delta_{i,1}+\delta_{i, 2n}),
\label{iir10}
\end{equation}
so the intensity vector becomes
\begin{eqnarray}
|I(\tau)\}&=&\sum_{i=1}^{2n}|v_i(\tau)\big)A_i\nonumber\\
&=&\frac{|v_1(\tau)\big)+|v_{2n}(\tau)\big)}{3\eta_1\tau_c}\nonumber\\
&=&|0).
\label{iir12}
\end{eqnarray}
One can derive the last line of (\ref{iir12}) from the previous line by inspection of (\ref{dd14}) and (\ref{dd16}).
For $\tau=0$ and $\tau=\tau_c$, the intracloud intensity vector  (\ref{iir12}) is consistent with (\ref{iir6}) and (\ref{iir8}).
For isotropic incident radiation, the intensity inside the cloud is also isotropic and has the same magnitude as the incident radiation.

We can think of the intracloud intensity vector (\ref{iir12}) as a part
$|I^{\{\bf u\}}(\tau)\}$ generated by the upward  part of the input intensity vector of (\ref{iir2}),  $\mathcal{M}_{\bf u}|I^{\{\rm in\}} \}=\mathcal{M}_{\bf u}|0)$, incident on the bottom of the cloud  and a part $|I^{\{\bf d\}}(\tau)\}$ generated by the downward input intensity vector $\mathcal{M}_{\bf d}|I^{\{\rm in\}} \}=\mathcal{M}_{\bf d}|0)$. incident on the top of the cloud. We use (\ref{iic2}) to write the intensity vector from the upward incoming intensity vector as
\begin{eqnarray}
|I^{\{\bf u\}}(\tau)\}&=&\sum_{i'i''}|v_{i'}(\tau)\big)\mathcal{I}^{-1}_{i'i''}\lvec\mu_{i''}|\mathcal{M}_{\rm u}|0) \nonumber\\
&=&\sum_{i'k}|v_{i'}(\tau)\big)\mathcal{I}^{-1}_{i'k}\lvec\mu_k|0) \nonumber\\
&=&\sum_{i'k}|v_{i'}(\tau)\big)\mathcal{I}^{-1}_{i'k}w_k. \nonumber\\
\label{iir14}
\end{eqnarray}
Here the first summation index $i'$ takes on the values $i'=1,2,3,\ldots, 2n$.
From the definition (\ref{sdbv9}) of $\mathcal{M}_{\bf u}$, we see that the second summation index takes on the values $i''=k=n+1,n+2,n+3,\ldots,2n$. From (\ref{sdbv4}) we see that $\lvec\mu_k|0) =w_k$. The projections $\lvec\mu_i|I^{\{\bf u\}}(\tau)\}$  are shown as the red rays in Figs. \ref{isin20} and \ref{isin1}.

In like manner we use (\ref{iic2}) to write the intensity vector from the downward incoming intensity as

\begin{eqnarray}
|I^{\{\bf d\}}(\tau)\}&=&\sum_{i'i''}|v_{i'}(\tau)\big)\mathcal{I}^{-1}_{i'i''}\lvec\mu_{i''}|\mathcal{M}_{\rm d}|0) \nonumber\\
&=&\sum_{i'j}|v_{i'}(\tau)\big)\mathcal{I}^{-1}_{i'j}w_j. \nonumber\\
\label{iir16}
\end{eqnarray}
Here the first summation index $i'$ takes on the values $i'=1,2,3,\ldots, 2n$.
From the definition (\ref{sdbv9}) of $\mathcal{M}_{\bf d}$, we see that the second summation index takes on the values $i''=j=1,2,3,\ldots,n$.
The projections $\lvec\mu_i|I^{\{\bf d\}}(\tau)\}$  are shown as the blue rays in Figs. \ref{isin20} and \ref{isin1}.

In summary, for the isotropic incoming intensity vector (\ref{iir2}), the intensities inside a cloud remain isotropic and equal to the input intensity at any optical depth $\tau$ above the bottom, for any scattering phase $p(\mu)$,  and for any cloud thickness $\tau_c$. For real absorbing clouds of sufficiently large optical thickness $\tau_c\gg 1$, the intensity attenuates to negligibly small values near the middle.

\section{Thin Clouds}
For infinitesimally thin clouds of optical depth $\tau_c=d\tau$ the scattering matrix simplifies to 
\begin{equation}
\mathcal{S}=\hat 1-d\tau (\hat\varsigma_{\bf u}-\hat\varsigma_{\bf d})\hat\eta.
\label{tcr2}
\end{equation}
in accordance with (WH-184).  The efficiency matrix $\hat\eta$ was given by (\ref{vet6}). From (\ref{c22a}) we find that the cloud albedo matrix becomes
\begin{eqnarray}
\Omega &=&(\hat\mu_{\bf u}-\hat\mu_{\bf d}) \mathcal{S}(\hat\varsigma_{\bf u}-\hat\varsigma_{\bf d})\nonumber\\
&=&\hat 1-d\tau \hat\eta(\hat\varsigma_{\bf u}-\hat\varsigma_{\bf d}).
\label{tcr4}
\end{eqnarray}
From (\ref{alb10dd}) we can write the absorption probability for the upward part of the incoming radiation as
\begin{eqnarray}
A_{\bf u}&=&1-\frac{\lvec 0|\{\hat 1-d\tau \hat\eta(\hat\varsigma_{\bf u}-\hat\varsigma_{\bf d})\}\mathcal{M}_{\bf u}|Z^{\{\rm in\}}\}}{\lvec 0|\mathcal{M}_{\bf u}|Z^{\{\rm in\}}\}}\nonumber\\
&=&d\tau\frac{\lvec 0| \hat\eta\hat\varsigma_{\bf u}|Z^{\{\rm in\}}\}}{\lvec 0|\mathcal{M}_{\bf u}|Z^{\{\rm in\}}\}}\nonumber\\
&=&d\tau (1-\tilde\omega)\frac{\lvec 0|\hat\varsigma_{\bf u}|Z^{\{\rm in\}}\}}{\lvec 0|\mathcal{M}_{\bf u}|Z^{\{\rm in\}}\}}
\label{tcr6}
\end{eqnarray}
Here we noted from (\ref{in12}) that $\lvec 0|\hat\eta=(1-\tilde\omega)\lvec 0|$.
From (\ref{tcr6}) we see that the absorption $A_{\bf u}$ of a thin cloud depends on the direction of the input flux, as described by $|Z^{\{\rm in\}}\}$, but is independent of the scattering phase function.

As a simple example, suppose that the incoming flux vector consists of a single upward stream of index $k>n$,
\begin{equation}
|Z^{\{\rm in\}}\}=4\pi\mu_k|\mu_k)
\label{tcr8}
\end{equation}
 Substituting (\ref{tcr8}) into (\ref{tcr6}) we find that the absorption is
\begin{eqnarray}
A_{\bf u}
&=&d\tau(1-\tilde\omega)\frac{\lvec 0| \hat\varsigma_{\bf u}|\mu_k)}{\lvec 0|\mathcal{M}_{\bf u}|\mu_k)}\nonumber\\
&=&d\tau\varsigma_k(1-\tilde\omega)\frac{\lvec 0|\mu_k)}{\lvec 0|\mu_k)}\nonumber\\
&=&d\tau\varsigma_k(1-\tilde\omega).
\label{tcr10}
\end{eqnarray}
According to (\ref{tcr10}), for an input intensity vector $|I^{\{\rm in\}}\}=|\mu_k)$, corresponding to the input flux vector (\ref{tcr8}), the absorption $A_{\bf u}$ is proportional to $d\tau\varsigma_k$, the slant distance of the input stream through the cloud, and to the probability $(1-\tilde\omega)$ that a photon collision with a cloud particulate leads to absorption rather than scattering.

As another example, suppose that the incoming intensity vector at the bottom of the cloud is half isotropic, $|I^{\{\rm in\}}\}=\mathcal{M}_{\bf u}|0)$, so that the incoming flux vector (\ref{alb0g}) becomes
\begin{equation}
|Z^{\{\rm in\}}\}=4\pi\hat\mu_{\bf u}|0).
\label{tcr12}
\end{equation}
Substituting (\ref{tcr12}) into (\ref{tcr6}) and noting that 
$\lvec 0|\mathcal{M}_{\bf u}|0)=1/2$, as follows from  (\ref{tcr34}) below, we find
\begin{eqnarray}
A_{\bf u}
&=&d\tau(1-\tilde\omega)\frac{\lvec 0|\mathcal{M}_{\bf u}|0)}{\lvec 0|\hat\mu_{\bf u}|\mu_k)}\nonumber\\
&=&\frac{d\tau(1-\tilde\omega)}{2\lvec 0|\hat\mu_{\bf u}|0)}
\label{tcr14}
\end{eqnarray}
The matrix element in the denominator of (\ref{tcr14}) can be written as
\begin{eqnarray}
\lvec 0|\hat\mu_{\bf u}|0)&=&\frac{1}{2}\sum_{k=n+1}^{2n}w_k\mu_k\nonumber\\
&\to&\frac{1}{2}\int_0^1d\mu \,\mu\nonumber\\
&=&\frac{1}{4},\quad\hbox{as}\quad n \to \infty.
\label{tcr16}
\end{eqnarray}
Here we made use of the limit of the Gauss-Legendre quadrature,
\begin{equation}
\lim_{n\to\infty}\sum_{i=1}^{2n} w_i f(\mu_i)=\int_{-1}^1 d\mu f(\mu).
\label{tcr18}
\end{equation}
Substituting (\ref{tcr16}) into (\ref{tcr14}) we find
\begin{equation}
\lim_{n\to\infty}A_{\bf u}=2\,d\tau  (1-\tilde\omega).
\label{tcr20}
\end{equation}
For quasi-isotropic illumination from one side, the effective thickness of a thin cloud is $2 d\tau$. 

\begin{figure}[t]
\postscriptscale{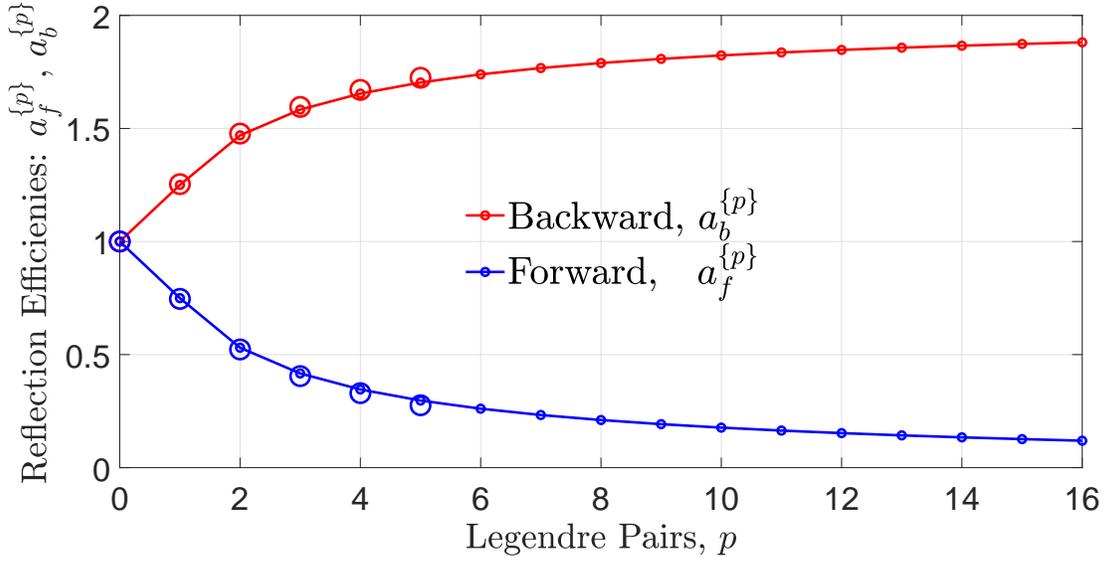}{1}
\caption{Reflection efficiencies for $2n=32$ streams of half-isotropic radiation (\ref{tcr12}), incident onto the bottom of a thin cloud. The small blue circles,  $a^{\{p\}}_f$, given by  (\ref{tcr44}) are for the maximum forward scattering phase function $p(\mu) = \varpi^{\{p\}}(\mu)$ of  
(\ref{pfb2}). The small red circles, given by (\ref{tcr46}),  are for the  the maximum backward scattering phase function  $p(\mu) = \varpi^{\{p\}}(-\mu)$. The values of $a^{\{p\}}_f=a^{\{p\}}_b=1$ for $p=0$ are for isotropic scattering.  As one would expect, strongly peaked forward-scattering phase functions, constructed from many pairs $p$ of Legendre polynomials, suppress the thin cloud reflection efficiency $a^{\{p\}}_f$, compared to isotropic scattering. Strongly backward-peaked phase functions can nearly double the reflection efficiency $a^{\{p\}}_b$. For $2n = 10$ streams the coefficients $a^{\{p\}}_f$  and $a^{\{p\}}_b$, shown as large circles, are almost the same as for $2n = 32$ streams. The largest difference is for $p=5$, where for $2n = 10$ streams we find
$a_f^{\{5\}} =0.2759 $ and for $2n = 32$ streams we find $a_f^{\{5\}} =0.2968 $.
\label{anp}}
\end{figure}

From (\ref{alb10f}) we see that the reflected fraction of the half-isotropic input flux vector (\ref{tcr12}) is
\begin{eqnarray}
R_{\bf u}&=&\frac{\lvec 0|\mathcal{M}_{\bf d}\Omega\hat\mu_{\bf u}|0)}{\lvec 0|\hat\mu_{\bf u}|0)}\nonumber\\
&=&\frac{\lvec 0|\mathcal{M}_{\bf d}(\hat\mu_{\bf u} -d\tau\hat\eta \mathcal{M}_{\bf u})|0)}{\lvec 0|\hat\mu_{\bf u}|0)}\nonumber\\
&=&-d\tau\frac{\lvec 0|\mathcal{M}_{\bf d}\hat\eta \mathcal{M}_{\bf u}|0)}{\lvec 0|\hat\mu_{\bf u}|0)}\nonumber\\
&=&-d\tau\frac{\lvec 0|\mathcal{M}_{\bf d}(\hat 1-\tilde\omega\hat p/2) \mathcal{M}_{\bf u}|0)}{\lvec 0|\hat\mu_{\bf u}|0)}\nonumber\\
&=&d\tau\frac{\tilde\omega\lvec 0|\mathcal{M}_{\bf d}\hat p\mathcal{M}_{\bf u}|0)}{2\lvec 0|\hat\mu_{\bf u}|0)}\nonumber\\
&=&d\tau\tilde\omega\sum_{l=0}^{2n-1}p_l\frac{\lvec 0|\mathcal{M}_{\bf d}|l)\lvec l|\mathcal{M}_{\bf u}|0)}{\lvec 0|\hat\mu_{\bf u}|0)}.
\label{tcr22}
\end{eqnarray}
To get the second line of (\ref{tcr22}) from the first we used (\ref{tcr4}) and the fact that $(\hat\varsigma_{\bf u}-\hat\varsigma_{\bf d}) \hat\mu_{\bf u}=\mathcal{M}_{\bf u}$. The third line follows from the second since $\mathcal{M}_{\bf d}\hat\mu_{\bf u} = \hat 0$. To get the fourth line from the third we used the expression (\ref{vet6}) for the efficiency matrix $\hat \eta$. To get the fifth line from the fourth, we noted from (\ref{sdbv9c}) that $\mathcal{M}_{\bf d}\mathcal{M}_{\bf u} =\hat 0$. To get the last line from the fifth, we used the multipole expansion (\ref{me12}) of the phase matrix $\hat p$.

Using (\ref{sdbv2}) and (\ref{sdbv4})  with (\ref{sdbv9}) we write 
\begin{eqnarray}
\lvec l|\mathcal{M}_{\bf u}|0)&=&\sum_{k=n+1}^{2n}\lvec l|\mu_k)\lvec\mu_k|0)\nonumber\\
&=&\frac{1}{2}\sum_{k=n+1}^{2n}w_kP_l(\mu_k).
\label{tcr24}
\end{eqnarray}
In like manner we write 
\begin{eqnarray}
\lvec 0|\mathcal{M}_{\bf d}|l)&=&\sum_{j=1}^n\lvec 0|\mu_j)\lvec\mu_j|l)\nonumber\\
&=&\frac{2l+1}{2}\sum_{j=1}^{n}w_jP_l(\mu_j)\nonumber\\
&=&(-1)^l\frac{2l+1}{2}\sum_{k=n+1}^{2n}w_kP_l(\mu_k)\nonumber\\
&=&(-1)^l(2l+1)\lvec l|\mathcal{M}_{\bf u}|0).
\label{tcr26}
\end{eqnarray}
For reflection-conjugate pairs of indices $j$ and $k$, with $k=2n+1-j$, we noted the symmetries 
$\mu_k=-\mu_j$,  $w_j = w_k$ and $P_l(\mu_j)=(-1)^lP_l(\mu_k)$, to get the third line from the second.
We used (\ref{tcr24}) to write the last line of (\ref{tcr26}).

 Using (\ref{tcr26}), we write (\ref{tcr22}) as
\begin{eqnarray}
R_{\bf u}
&=&\frac{d\tau\tilde\omega}{\lvec 0|\hat\mu_{\bf u}|0)}
\sum_{l=0}^{2n-1}(-1)^l(2l+1)p_l\lvec l|\mathcal{M}_{\bf u}|0)^2.
\label{tcr28}
\end{eqnarray}
Using the symmetries mentioned  above, we see from (\ref{tcr24}) that
\begin{equation}
\lvec l|\mathcal{M}_{\bf u}|0)=(-1)^l\lvec l|\mathcal{M}_{\bf d}|0).
\label{tcr30}
\end{equation}
Using (\ref{in12cc}), (\ref{sdbv9a}) and (\ref{tcr30}) we find
\begin{eqnarray}
\delta_{l0}&=&\lvec l|0)=\lvec l|\hat 1|0)\nonumber\\
&=&\lvec l|\mathcal{M}_{\bf u}+\mathcal{M}_{\bf d}|0)\nonumber\\
&=&\lvec l|\mathcal{M}_{\bf u}|0)[1+(-1)^l].
\label{tcr32}
\end{eqnarray}
For even values of the multipole index $l$ we can write (\ref{tcr32})
\begin{equation}
\lvec l|\mathcal{M}_{\bf u}|0)=\frac{\delta_{l0}}{2},\quad\hbox{for}\quad l=0,2,4,\ldots.
\label{tcr34}
\end{equation}
We can therefore write the reflectivity (\ref{tcr28}) as the product of a reflectivity $R_{\bf u}^{\{\rm iso\}}$ of a thin cloud with isotropic scattering and an anisotropic scattering efficiency $a$,
\begin{equation}
R_{\bf u}=R_{\bf u}^{\{\rm iso\}}a.
\label{tcr38}
\end{equation}
Using (\ref{tcr14}), we find that
the reflectivity for isotropic scattering, with $p_l=\delta_{l0}$, is
\begin{equation}
R_{\bf u}^{\{\rm iso\}}=\frac{d\tau\tilde\omega
}{4\lvec 0|\hat\mu_{\bf u}|0)}=\frac{\tilde\omega}{2(1-\tilde\omega)}A_{\bf u}\approx \tilde\omega d\tau
\label{tcr40}
\end{equation}
The anisotropic scattering efficiency  is
\begin{eqnarray}
a=
1-4\sum_{l'}(2l'+1)p_{l'}\lvec l'|\mathcal{M}_{\bf u}|0)^2,\quad
\hbox{where}\quad
l' = 1,3,5,\ldots,2n-1.
\label{tcr42}
\end{eqnarray}
Formulas for downward radiation can be obtained by replacing the index $\{\bf u\}$ by $\{\bf d\}$ in (\ref{tcr38})--(\ref{tcr42}).

Examples of (\ref{tcr38}) are shown in Fig. \ref{anp} for the maximum forward-scattering phase functions, $p(\mu) = \varpi^{\{p\}}(\mu)$ of 
(\ref{pfb2}). We denote the anisotropic efficiencies for maximum forward scattering  by
\begin{eqnarray}
a^{\{p\}}_f=
1-4\sum_{l'}(2l'+1)\varpi^{\{p\}}_{l'}\lvec l'|\mathcal{M}_{\bf u}|0)^2,\quad
\hbox{where}\quad
l' = 1,3,5,\ldots,2n-1.
\label{tcr44}
\end{eqnarray}
The anisotropic efficiencies for maximum backward scattering, where the scattering phase is $p(\mu) = \varpi^{\{p\}}(-\mu)$,  are 
\begin{eqnarray}
a^{\{p\}}_b=
1+4\sum_{l'}(2l'+1)\varpi^{\{p\}}_{l'}\lvec l'|\mathcal{M}_{\bf u}|0)^2,\quad
\hbox{where}\quad
l' = 1,3,5,\ldots,2n-1.
\label{tcr46}
\end{eqnarray}
We see that 
\begin{eqnarray}
a^{\{p\}}_f+a^{\{p\}}_b = 2.
\label{tcr48}
\end{eqnarray}
As one can see from Fig. \ref{anp}, the  numerical coefficients $a^{\{p\}}_f$ and  $a^{\{p\}}_b$  are nearly the same for radiation transfer represented by $n=5$ stream pairs, shown as large circles, and $n=16$ stream pairs, shown as small circles.  The number of Legendre-polynomial pairs $p$ in the phase functions $\varpi^{\{p\}}(\mu)$ cannot exceed $n$, that is $p\le n$.

\section{Summary}
We  have shown how to use matrix methods of quantum mechanics to efficiently and accurately calculate axially symmetric  radiation transfer in clouds, with conservative multiple scattering, and with arbitrary single-scattering anisotropies or phase functions.

At a given optical thickness, $\tau$, we model the radiation with the $2n$ unweighted intensities $I(\mu_i,\tau)$ of (\ref{in7}), which are values of the intensity $I(\mu,\tau)$ at the Gauss-Legendre direction cosines  $\mu = \mu_i$.  The $\mu_i$ are the roots of the Legendre polynomial $P_{2n}(\mu)$.  According to  (\ref{in6}), the $\mu_i$ are also the eigenvalues of the $2n\times 2n$ direction-cosine matrix $\hat\mu$ of (\ref{me3}). As shown in (\ref{in7}), it is convenient to write the unweighted intensities as
$I(\mu_i,\tau)=w_i^{-1}\lvec\mu_i|I(\tau)\}$ where the Gauss-Legendre weights can be defined by 
(\ref{in7a}).  Here the weighted intensity $\lvec\mu_i|I(\tau)\}$ denotes the projection of  an abstract intensity vector $|I(\tau)\}$ onto the left stream vector $\lvec \mu_i|$ of (\ref{in6}).

For clouds with non-zero absorption and $\tilde\omega <1$, WH showed how to use matrix methods to model $2n$-stream radiative transfer  with arbitrary scattering anisotropy. In WH the intensity vector $|I(\tau)\}$ of (\ref{in2}) was expanded onto $2n$ penetration-modes $|\lambda_i)$ which depend on optical depth $\tau$ as  $e^{-\tau/\lambda_i}$. Here $\lambda_i$ is the penetration length and $|\lambda_i)$ is the right eigenvector of the penetration-length matrix $\hat\lambda=\hat\kappa^{-1}$ the inverse of the exponentiation-rate matrix  $\hat \kappa$ of (\ref{dd2a}). When  $\tilde\omega \to 1$, and absorption vanishes entirely, two of the penetration lengths diverge to infinity, $\lambda_1\to -\infty$ and $\lambda_{2n}\to\infty$. Then it is necessary to replace the two infinitely-penetrating modes $|\lambda_1)$ and $|\lambda_{2n})$ with the quasi-isotropic bases of $|v_1\big)$  of (\ref{dd14}) and $|v_{2n})$ of (\ref{dd16}) that depend linearly, not exponentially on $\tau$. 
For $i=2,3,4,\ldots 2n-1$ the directional bases $|v_i)$ are the limits of the penetration modes of WH for $\tilde\omega \to 1$, that is $|\lambda_i)\to |v_i(\tau_i)\big) $, where the reference optical depths $\tau_i$ were given by (\ref{dd42}).  Quasi-isotropic and directional bases are illustrated in Fig. \ref{muv2} and \ref{muv3}.  The dependence of the penetration lengths $\lambda_i$  on the sharpness of the forward scattering phase function $\varpi^{\{p\}}(\mu)$  of (\ref{pfb2}) is shown in Fig. \ref{lamip}.

For conservative scattering, the flux $Z$ and the proportional dipole moment $I_1$ of the intensity are independent of optical depth $\tau$ in the cloud, as discussed  in connection with (\ref{in14c}).  Also, the $K$-integral,  or the second moment $K$ of the intensity, given by (\ref{in5bb}), varies linearly with $\tau$, as shown by (\ref{ps16}).  The invariance of $Z$ and the linear variation of $K$ with $\tau$ are illustrated with the moment diagrams of Figs. \ref{mom2t20v}, \ref{mom3t20v}, \ref{mom3t20h} and \ref{mom3t20i}.  

As directional radiation penetrates a cloud, multiple conservative scattering causes the radiation to isotropize within a few optical depths of the input surface. The isotropization is illustrated with the ray diagrams of Figs. \ref{ctr2t20v}, \ref{ctr3t20v}, and \ref{ctr3t20h}. The Eddington criterion (\ref{dd53}), that the higher multipole moments, $I_2(\tau), I_3(\tau), I_4(\tau),\ldots$, are negligible compared to the first two moments, $I_0(\tau)$ and $I_1(\tau)$, is very well satisfied near the centers of optically thick clouds, as one can see from the dashed red curves of  Figs. \ref{mom2t20v}, \ref{mom3t20v}, \ref{mom3t20h} and \ref{mom3t20i}.

The transmission and reflection of clouds depends strongly on the angle of incidence of incoming radiation. For the same thickness, $\tau_c = 20$,  and phase function, $p(\mu) =\varpi^{\{5\}}(\mu)$, a cloud transmits about three times more  near-normal-incidence radiation, shown in Fig. \ref{mom3t20v} or Fig. \ref{ctr3t20v}, than near-horizontal-incidence radiation of Fig. \ref{ctr3t20h} or Fig. \ref{mom3t20h}.  Fig. \ref{Ref2} shows how the reflectance of a cloud with the Rayleigh-scattering phase function of (\ref{rs2})  depends on the direction cosine $\mu_i$ of the input radiation and on the  cloud thickness $\tau_c$. Fig. \ref{Ref3} shows that a cloud with the more strongly forward scattering phase function $p(\mu)=\varpi^{\{5\}}(\mu)$ of (\ref{pfb2})  reflects less (and transmits more).
 
For a cloud illuminated from the bottom with equal intensity from all upward directions and illuminated from the top with the same equal intensity from all downward directions, the intensity that results from multiple scattering is isotropic and of equal intensity inside and outside the cloud. This ``conservation of isotropy" is independent of the scattering phase function as is illustrated in Fig. \ref{isin20} for forward scattering in a cloud with a relatively large optical depth, $\tau_c = 20$  and in Fig. \ref{isin1} for a thinner cloud with $\tau_c = 1$. Even for $\tau_c\to \infty$, the cloud will remain filled with isotropic radiation with equal intensities at all depths.  Real clouds of sufficiently large optical depth $\tau_c$, and with finite absorption, $1-\tilde\omega>0$ have negligible intensity at their centers, $\tau = \tau_c/2$, under the same conditions of half-isotropic illumination from the top and bottom.
The input radiation from both the top and bottom is absorbed before reaching the center. 

Fig. \ref{anp} shows how the reflectance of a thin cloud with optical thickess $d\tau\ll 1$ decreases with  increasing numbers $p$ of Legendre-polynomial pairs in the maximum forward-scattering phase functions $p(\mu)=\varpi^{\{p\}}(\mu)$ of (\ref{pfb2}).  The forward-scattering phases are $\varpi^{\{p\}}(1) =p(p+1)$, so the integer $p=1,2,3,\ldots$ measures the forward peaking of the phase functions.
Fig. \ref{anp} also shows that increasingly sharply peaked backscattering phase functions $p(\mu)=\varpi^{\{p\}}(-\mu) $  increase the reflectance toward a limiting value, $\mathcal{R}_{\bf u}\to 2\mathcal{R}^{\{\rm iso\}}_{\bf u}$ that is twice that of an isotropically scattering phase function, $\mathcal{R}^{\{\rm iso\}}_{\bf u}$, given by (\ref{tcr40}).

The number $2n$ of streams needed to make accurate calculations is often surprsisingly small. An example is shown in Figs. \ref{Ref2} and \ref{Ref3}, which show the reflectance of clouds versus the angle of incidence of input radiation. Calculations made with $2n = 32$ streams are indicated with small circles, and those made with $2n = 10$ streams are shown as large circles. The results are virtually the same.  Another example  is provided by Fig. \ref{anp}. For $p=5$, the maximum value possible for $n=5$, the results for $n=5$ and $n=16$ stream pairs differ by about 7\%.

Using the efficient matrix methods outlined above with the capabilities  of modern mathematical software makes calculations of radiation transfer  fast and simple.  The figures of this paper were generated with a few dozen lines of Matlab code.

\section*{Acknowledgements}
 The Canadian Natural Science and Engineering Research  Council provided financial support of one of us. We are grateful to Professor C. A. de Lange for valuable discussions of this work.

\end{document}